\documentclass[
twocolumn,
tightenlines,
10pt,
longbibliography,
showpacs,
nofootinbib,
notitlepage,
superscriptaddress]{revtex4-2}

\usepackage{amsmath, amssymb, amsthm, braket, enumitem, dsfont, relsize}
\usepackage{makecell, graphicx, subfiles, multirow, wrapfig, tabularx}
\usepackage[font=scriptsize]{subcaption}
\usepackage[font=scriptsize, justification=raggedright]{caption}
\usepackage[page]{appendix}
\usepackage[colorlinks, allcolors=black]{hyperref}
\usepackage[capitalize, nameinlink]{cleveref}
\usepackage{tikz}
\usetikzlibrary{quantikz}
\usetikzlibrary{shapes.geometric, arrows}
\usetikzlibrary{positioning}
\setlist[itemize]{leftmargin=10pt}

\DeclareCaptionLabelFormat{bold}{\textbf{(#2)}}
\begin{document}

\title{Mixed-Integer Programming Using a Bosonic Quantum Computer}

\author{Farhad~Khosravi}
\affiliation{1QB Information Technologies (1QBit), Vancouver, BC, Canada}

\author{Artur~Scherer}
\affiliation{1QBit, Waterloo, ON, Canada}

\author{Pooya~Ronagh}
\thanks{{\vskip-10pt}{\hskip-8pt}Corresponding author: \href{mailto:pooya.ronagh@1qbit.com}{pooya.ronagh@1qbit.com}\\}
\affiliation{1QB Information Technologies (1QBit), Vancouver, BC, Canada}
\affiliation{Institute for Quantum Computing, University of Waterloo, Waterloo, ON, Canada}
\affiliation{Department of Physics \& Astronomy, University of Waterloo, Waterloo, ON, Canada}
\affiliation{Perimeter Institute for Theoretical Physics, Waterloo, ON, Canada}

\date{\today}

\begin{abstract}
We propose a scheme for solving mixed-integer programming problems in which the
optimization problem is translated to a ground-state preparation problem on a set of bosonic quantum field modes (qumodes). We perform numerical demonstrations by simulating a circuit-based optical quantum computer with each
individual qumode prepared in a Gaussian state. We simulate an adiabatic evolution from an initial mixing Hamiltonian, written in terms of the momentum operators of the qumodes, to a final Hamiltonian which is a polynomial of the
position and boson number operators. In these demonstrations, we solve a variety
of small non-convex optimization problems in integer
programming, continuous non-convex optimization, and mixed-integer programming.
\end{abstract}

\maketitle

\section{Introduction}

Achieving a quantum advantage in solving mathematical optimization problems will
tremendously broaden the range of applications of quantum computers. To this
end, a great variety of approaches has been explored, from the experimental
realization of quantum annealers \cite{johnson2011quantum} to the design of circuit
model quantum algorithms for optimization \cite
{montanaro2020quantum, van2020quantum, alexandru2020quantum}. Quantum annealers
rely on the analog simulation of an adiabatic evolution process, which restricts the sizes of optimization problems they can solve in the limited time available before the system decoheres. Quantum circuit model
algorithms for optimization typically offer an asymptotic, yet modest,
quadratic speedup compared to the exponentially costly exhaustive search. This
quantum speedup relies on quantum amplitude amplification and estimation
techniques~\cite{brassard2000quantum}, and will likely require very large
fault-tolerant quantum computers~\cite{sankar2021benchmark} to solve problems of a size practical for real-world applications. In addition, both approaches
perform quantum algorithms on a register of qubits, which restricts the types of optimization problems they can solve, especially in the absence of dense couplings between the qubits.

In this paper, we use a set of bosonic quantum field modes (or ``qumodes''\footnote{The term 
``qumode'' is commonly used to refer 
to the quantum harmonic oscillator 
associated with each single mode of a bosonic quantum field. Unlike in the case of qubits,  
quantum information is encoded in the infinite-dimensional Fock space associated with 
each qumode.}) to encode a mixed-integer
programming (MIP) problem. We translate integer and continuous variables to
different observables of the bosonic system. Non-negative integer variables are
represented by the eigenstates of the number operators $\hat{n}= \hat{a}^\dagger \hat{a}$ of the
qumodes. Here, $\hat{a}^\dagger$ and $\hat{a}$ are, respectively,  the bosonic creation and
annihilation operators. In contrast, continuous variables are encoded in the
continuous spectrum of the quadrature operators $\hat{x}= \frac{1}{\sqrt 2}
(\hat{a} + \hat{a}^\dagger)$ or $\hat{p}= \frac{1}{i\sqrt 2}(\hat{a} - \hat{a}^\dagger)$, that is, the canonically conjugate position
and momentum observables of quantum harmonic oscillators, respectively.

Computation using bosonic systems has been explored in various
ways to date. Continuous-variable quantum computation (CVQC)~\cite{Lloyd1999quantum} pertains to performing unitary transformations according to
Hamiltonians that are polynomials of quadrature operators. In contrast, in
boson sampling~\cite{spring2013boson, Tillmann2013experimental,
ZHONG2019experimental}, quantum information is encoded in the discrete spectrum of the boson number operators. Both computing tasks are of experimental and theoretical interest in
the particular case of quantum computation using Gaussian states~\cite
{weedbrook2012gaussian, Bromley2020applications}. In addition, advances in
measurement-based, one-way universal quantum computation using CVQC~\cite
{Menicucci2006universal} and the creation of cluster states in optical
settings~\cite{Yokoyama2013ultra, Roslund2014wavelength} provide promising
prospects for experimental CVQC.

\vspace{1pt}\noindent{\bf Computational significance.}
A mathematical programming problem is an optimization problem in which an objective function is optimized with respect to a set of variables
satisfying a set of constraints. In MIP, both the
objective function and constraints are polynomials of integer and continuous
variables. Mixed-integer linear programming pertains to problems in which
the objective function and constraints are linear functions of the variables. This is enough to obtain natural models for typical combinatorial and discrete optimization problems~\cite
{sierksma2015linear}, including famous NP-hard optimization problems in
constraint satisfaction, graph and network optimization, and job scheduling~\cite{karp1972reducibility}. Mixed-integer nonlinear programming problems extend this class
to a boundless domain of real-world applications in science, engineering,
operations research, management, health care, decision making, and system control~\cite{floudas1995nonlinear, li2006nonlinear}. Despite this flexible
modelling power, classical algorithms for solving nonlinear and mixed-integer
optimization problems scale poorly with respect to problem size~\cite{burer2012non}. For instance, solving non-convex quadratic optimization
problems is itself extremely challenging, to the extent that even verifying 
local optimality of their feasible points is NP-hard~\cite{pardalos1988checking}.

In this paper, we recast an MIP problem as the problem of preparing the
ground state of a quantum Hamiltonian in a multi-mode Fock space, and then investigate
an adiabatic scheme for finding the ground state of the Hamiltonian. However,  other state preparation protocols, such as variational quantum algorithms~\cite{cerezo2021variational}, may be envisaged. Indeed, Verdon et al.~\cite
{verdon2019quantum} implement a continuous-variable analogue to the quantum
approximate optimization algorithm (QAOA)~\cite{farhi2014quantum} by writing
the problem Hamiltonian in terms of one quadrature operator and the mixing Hamiltonian in terms of the other.
However, to the best of our knowledge, the use of Fock states to encode
integer variables in the context of solving MIP problems, and the capability and computational significance of solving mixed-integer
non-convex global optimization problems using CVQC, have not been previously explored.

\vspace{1pt}\noindent{\bf Numerical demonstrations.}
In \cref{sec:ILP}, we investigate the optimization of integer linear programming~(ILP) problems, and solve a small example instance of the NP-hard unbounded
knapsack problem~\cite{lueker1975two}. In \cref{sec:NLIP}, we use a quadratic
binary formulation of the NP-hard \textsc{MaxClique} problem to show that our scheme works, without any substantial modifications, in the case of nonlinear
integer programming (NLIP) problems. Then, in~\cref{sec:Non-convex}, we use the connection between the \textsc{MaxClique} problem and continuous non-convex optimization, established by Motzkin and Straus~\cite{motzkin1965maxima}, to explore global non-convex
optimization via both continuous and integer implementations of the \textsc{MaxClique} problem. We solve the problem using only the quadrature operators in the case of the continuous formulation and only the number operators in the case of the integer formulation. Finally, in~\cref{sec:MIP}, we demonstrate the application of our method in solving MIP problems, using an example instance of a sparse optimization problem. Such cardinality-constrained
optimization problems are also NP-hard~\cite{bienstock1996computational} and of
practical importance in compressed sensing, signal processing, and
computational finance~\cite{tillmann2021cardinality}.

For these demonstrations, we simulate a circuit model optical device, where individual qumodes are prepared as Gaussian states. An adiabatic evolution process is
executed beginning with Gaussian states, followed by homodyne or photon number resolving
 measurements of the output qumodes, with the goal of preparing a
state that overlaps substantially with the ground state of a problem Hamiltonian.

\vspace{1pt}\noindent{\bf Experimental considerations.}
In a gate-based optical device, qumodes can be prepared as Gaussian states, after which they are acted upon by single- or two-qumode quantum optical gates integrated in photonic
or optical fibre waveguides~\cite{O'Brien2009photonic, Sparrow2018simulating,
Arrazola2021quantum}. While suffering from their own limitations, quantum
optical devices have an advantage in that they can operate at room temperature
and are easier to scale compared to other implementations of quantum
processors~\cite{Rudolph2017why}.

Current realizations of experiments that implement photon--photon interactions restrict the practical realization of our method to solving integer programming problems that are at most
quadratic in terms of integer variables. The computation is carried out by a set of
gates implementing functions of the photon number operator, including the
rotation gate $\hat{R}(\phi)=\exp(i\phi\hat{n})$, the Kerr gate $\hat{K}
(\alpha)=\exp(i\alpha\hat{n}^{2})$, and the cross-Kerr gate $\hat{CK}
(\alpha)=\exp (i\alpha\hat{n}_{1}\hat{n}_{2})$, the last of which acts on the Fock space of two
qumodes~\cite{Killoran2019strawberryfields}. Together with the quadratic phase
gate $\hat{P} (\alpha)=\exp(i\alpha\hat{p}^{2})$ and the position displacement
gate $\hat{X} (x)=\exp(ix\hat{p})$, this gate set is sufficient for solving
integer linear and quadratic programming problems. To solve MIP problems, we
require combined unitary gates that are functions of $\hat{n}$ and one of the quadrature operators; for example, $\hat{XN}(\alpha)=\exp(i\alpha\hat{x}^2_{1}\hat{n}_
{2})$. While unitary operations that use such a combination of quadrature and photon number operators have not been  realized,  we believe
that our results will provide the impetus for the experimental realization of such unitary operations. Toward the realization of a universal gate set for CVQC, Yanagimoto et al.~\cite{yanagimoto2020engineering} propose a promising deterministic implementation of the non-Gaussian cubic phase gate $\exp\left({i\alpha \hat{x}^3}\right)$, an approach worth exploring for implementing the two-qumode higher-order combined gates we have introduced. 

Implementing non-Gaussian gates at a high level of fidelity remains an experimental
challenge. While on-chip \mbox{Mach--Zehnder} interferometers based on beam
splitters and thermo-optic phase shifters are efficiently fabricated for high-fidelity performance~\cite{Clements2016integrated, Harris2017quantum,
Wang2018multidimensional}, the viability of creating photon--photon
interactions, such as in  Kerr and cross-Kerr gates, through $\chi^{(3)}$
third-order optical nonlinearity, has been subject to skepticism due to various
challenges hampering the performance of these \mbox{gates~\cite{Chuang1995simple,
Shapiro2006single, Gea2010impossibility, continuous2012bing}}. However,
ongoing research on photon--photon interactions points to the possibility generating phase shifts from single-photon pulses using other methods such as
quantum dots~\cite{Fushman2008controlled}, Rydberg blockades~\cite{Tiarks2019photon, Hamsen2018strong}, and four-wave mixing using atomic
systems~\cite{sagona2020conditional}. These experiments show the physical
possibility of creating Kerr interactions through a variety of approaches. The results
presented in this paper motivate experimental efforts toward the realization of quantum unitary operations that incorporate \mbox{photon--photon} interactions. Here, we rely on the assumption of the existence of
a viable method of implementing high-fidelity Kerr and cross-Kerr interactions
(regardless of the method of their realization), which are necessary to implement the quantum
operations required for MIP using CVQC.

\section{Adiabatic ground-state preparation}
\label{sec:QAO}

We study the probability distribution of states prepared via
adiabatic quantum computation (AQC). The time-dependent Hamiltonian considered
for AQC is
\begin{equation}
\hat{H}(\tau)= (1-\tau)\hat H_\text{M} + \tau \hat H_\text{P},
\label{eq:lin-adiabatic-schedule}
\end{equation}
where $\tau=t/T$ is the normalized time and $T$ is the total evolution time. The Hamiltonian $\hat H_\text{P}$ represents the \emph{problem} Hamiltonian
whose ground states encode the solutions to the optimization problem of
interest, and $\hat{H}_\text{M}$ is called a \emph{mixing} Hamiltonian and is required
to have an easy-to-prepare ground state and not to commute with~$\hat H_\text{P}$.

When $\hat H_\text{M}$ and $\hat H_\text{P}$ have discrete spectra, the evolution time $T$ is in $\mathcal{O}(\delta^{-2})$, where $\delta$ is the smallest spectral gap of $\hat H_\text{P}$ for all $\tau$~\cite{farhi2000quantum}. Therefore,  implementing this evolution in a scalable way via analog computation can be challenging~\cite{crosson2021prospects}. While for AQC involving Hamiltonians with continuous spectra the evolution time cannot be bounded in terms of a well-defined spectral gap, results similar to the conventional adiabatic theorem~\cite{maamache2008adiabatic} suggest that the evolution in time $T$ deviates from identity by $\mathcal{O}(1/T)$. However, the evolution can be discretized and implemented on a fault-tolerant circuit model quantum device. 
To this end, we use the common technique of Trotterization 
to approximate the adiabatic time evolution.
The resulting circuit consists of iterative applications
of unitary evolutions according to $\hat H_{\text{M}}$ and $\hat H_{\text{P}}$,
\begin{equation}\hat{U} =
e^{-ib_k\hat H_{\text{M}}}e^{-ic_k\hat H_{\text{P}}}\cdots
e^{-ib_1\hat H_{\text{M}}}e^{-ic_1\hat H_{\text{P}}}\,,
\label{eq:Trotterization}
\end{equation}
to an initial state $\ket{\psi_0}$ prepared in the ground state
of $\hat{H}_{\text{M}}$. The overall number of Trotter steps, denoted by $k$, directly pertains to the circuit depth of the algorithm. The choice of the coefficients in the exponents   
depends on the function describing the schedule. For a linear schedule as given in \cref{eq:lin-adiabatic-schedule}, 
$b_j=\left(k-j + \frac{1}{2}\right) \frac{T}{k^2}$ and 
$c_j=\left(j - \frac{1}{2}\right) \frac{T}{k^2}$. 
Note that the sum of the coefficients of both the mixing and problem Hamiltonians is equal to $T$, and that their coefficients decrease and increase linearly with each Trotter step $j$, respectively. In our numerical simulations, we find that the continuous AQC algorithm and its discretized version yield very similar final results for all of the problems studied. For this reason, except in the case of the feasibility problem (see \cref{sec:ILP}), we present the results only for the continuous algorithm. 

The mixing Hamiltonians used in our work are of the form
\begin{equation}
\hat{H}_\text{M}=\sum_{i=1}^{N}\left(\hat{p}_{i}^{2}-2p_{0i}\hat{p}_{i}\right), \label{eq:Mixing}
\end{equation}
which is equivalent to $\sum_{i=1}^N (\hat{p}_{i}-p_{0i})^2$
whose ground state is the momentum eigenstate
\mbox{$\ket{p_0}= \ket{p_{01}}\otimes \cdots \otimes \ket{p_{0N}}$.} Here, $N$ is the
number of qumodes. The quantum state $\ket{p_0}$ is a negatively and infinitely
squeezed coherent state along the momentum axis~\cite{gerry2005introductory}.
Although such states are not physically realizable, we can approximate them with the
squeezed coherent states $\ket{\mathbf{\boldsymbol{\alpha}},-{\boldsymbol{r}}}$, where
\mbox{$\boldsymbol{\alpha}=(\alpha_1,\alpha_2, \ldots, \alpha_N)$} and \mbox{$\boldsymbol{r}=(r_1,r_2, \ldots, r_N)$.} In all our simulations, \mbox{$\alpha_{j}=ip_{0j}/\sqrt{2\hbar}$}
and $r_{j}=r$ for all $j$, where $r$ is an experimentally feasible squeezing
parameter, for example, taking values between $0.5$ and $0.8$.
As the values of $p_{0i}$ increase, these states overlap increasingly less with the vacuum state. This is important, as the vacuum state is an eigenstate of the
number operator that is used to construct the problem Hamiltonian, and having a large
overlap between the initial state and the eigenstates of the problem Hamiltonian
will impair the adiabatic ground-state preparation.

In general, the initial state parameter $p_0$ affects the final quantum
states obtained via the adiabatic scheme and can be treated as a hyperparameter
that gives us direct access to a continuum of the initial mixing Hamiltonians
and the associated ground states to use in our algorithm. For example, if $\hat H_\text{P}$
has degenerate ground states, $p_0$ can be tuned to amplify the amplitude of each of those states or, alternatively, to create a uniform superposition of all of those states~(see \cref{sec:fair_sampling}). The latter is known as \emph{fair sampling}, which is 
equivalent to Gibbs sampling at the zero temperature limit and is therefore a
desirable property in machine learning~\cite{crawford2016reinforcement,
sepehry2018smooth}. In addition, for a generic problem Hamiltonian written in
terms of $\hat n $ and $\hat x$ operators, the mixing Hamiltonian
above satisfies the non-commutativity relation $[\hat H_\text{P},\hat H_\text{M}]\neq0$.

\section{Integer Linear Programming}
\label{sec:ILP}

\begin{figure*}[t]
\subfloat[\label{fig:IP_adiabatic_final_states}]
{\includegraphics[scale=0.25]{figs/probs_ILP_adiabatic_steps9001_time50_p0.72}}
\hspace{2mm}
\subfloat[\label{fig:IP_adiabatic_evolution}]
{\includegraphics[scale=0.25]{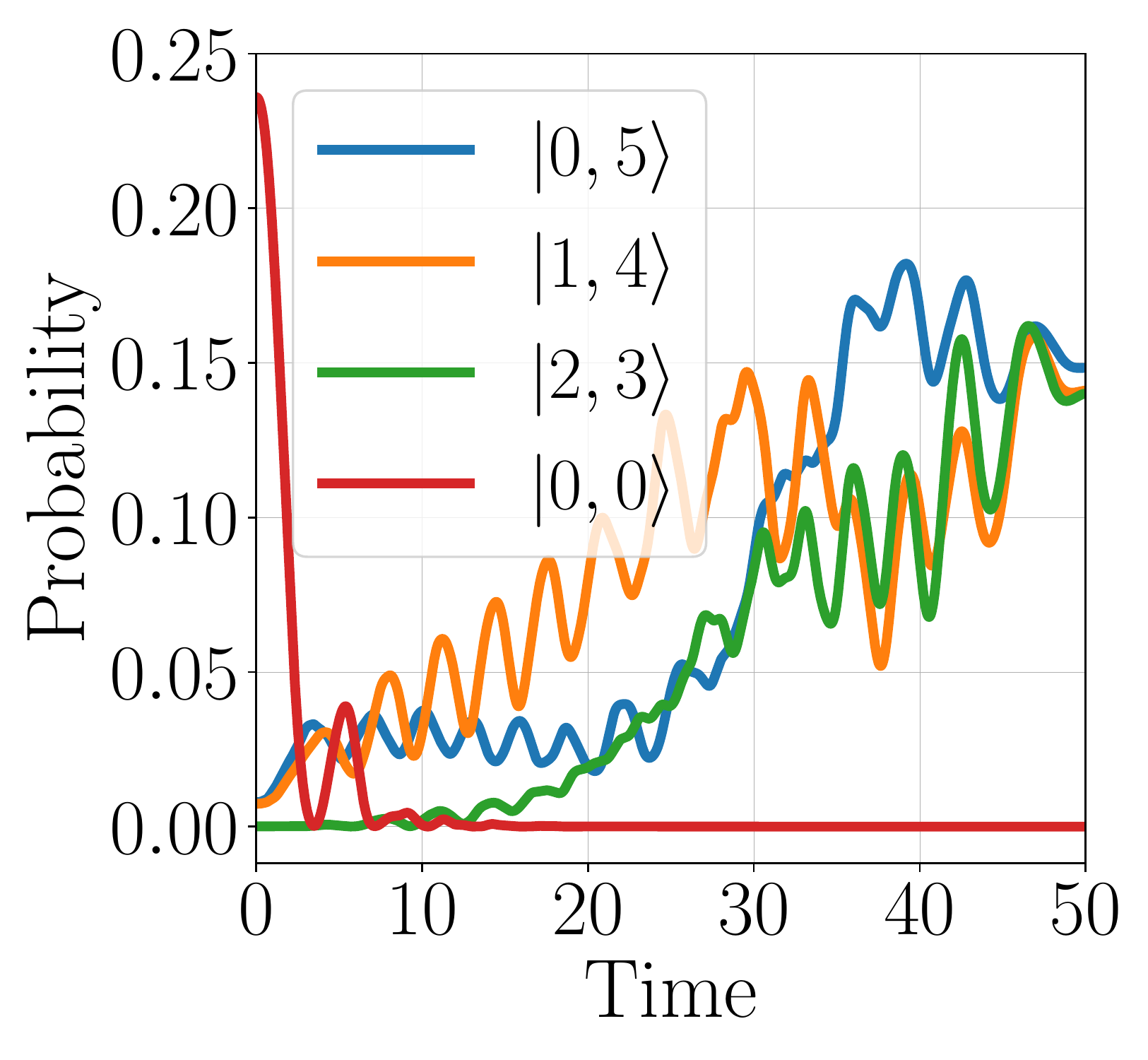}}
\hspace{2mm}
\subfloat[\label{fig:IP_QAOA_evolution}]
{\includegraphics[scale=0.25]{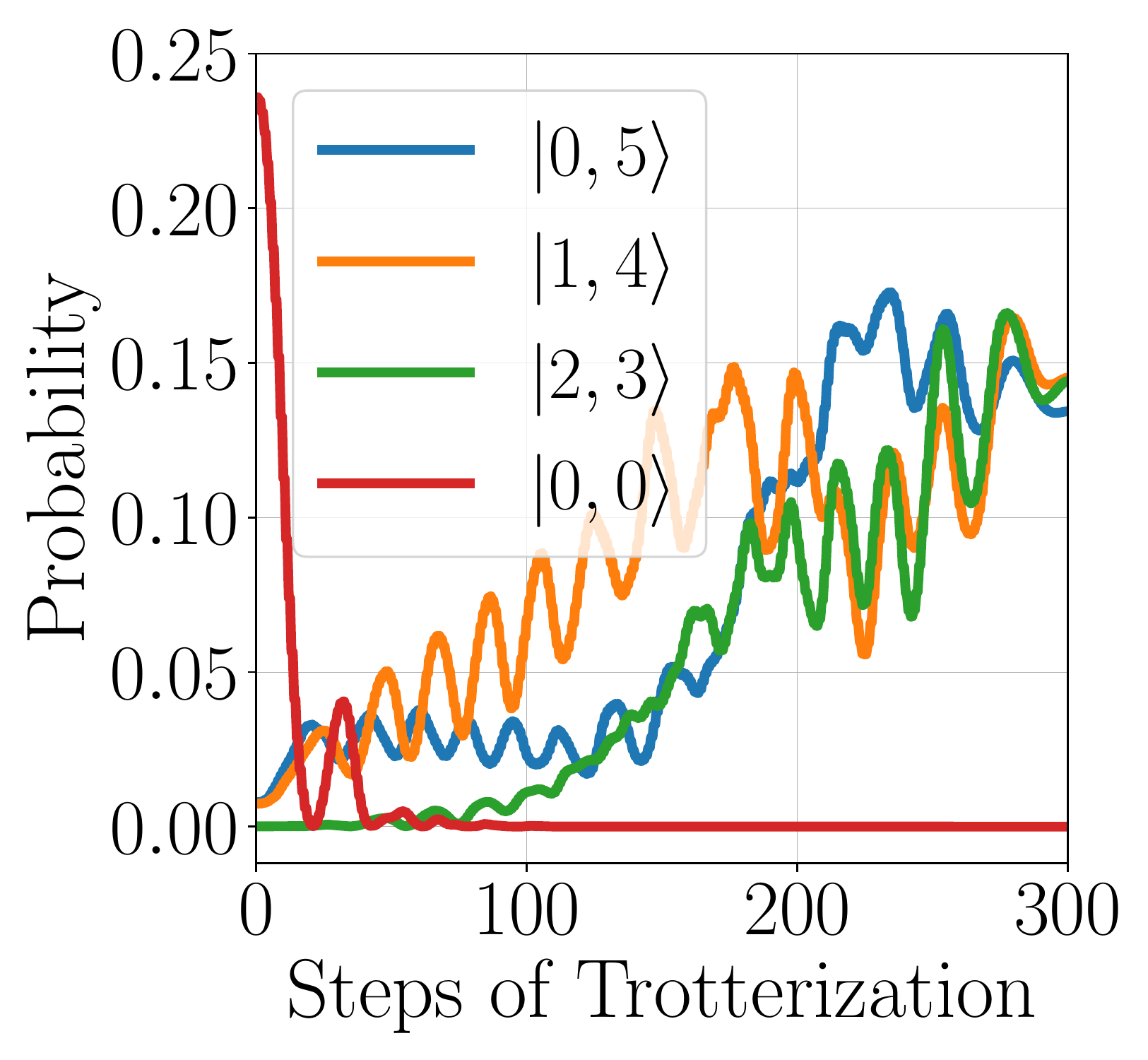}}

\caption{Results for the feasibility problem specified by the equation $n_1 + n_2= 5$, solved using AQC and its discretized form. (a) The probability
distribution of the final quantum states in the two-qumode Fock basis
$\{\ket{n_1, n_2}\}$. (b) The evolution of the probabilities of the dominant Fock states (in this case those that have the highest final probability amplitudes)  during AQC for the total evolution time $T= 50$. (c) The evolution of the probabilities of the dominant Fock states during the execution of discretized AQC  for a total number of Trotter steps $k= 300$. (b) and (c) The qumodes are initialized
as the squeezed coherent states $\ket{\mathbf{\boldsymbol{\alpha}}, {-\boldsymbol{r}}}$, where
$\boldsymbol{\alpha}=(\alpha_1,\alpha_2)$, with $\alpha_1 = \alpha_2= 0.72i/\sqrt{2}$and $r =0.8$. Due to the symmetry between ${n_1}$ and ${n_2}$, if the Fock state $\ket{n_1, n_2}$ is a solution, then $\ket{n_2, n_1}$ is also a solution. For this reason, only three of the six degenerate ground states are shown.}
\label{fig:evolutions_IP}
\end{figure*}

Every ILP problem can be written in the \emph{canonical} form
\begin{equation}
\begin{split}
\text{minimize\quad} & -c^T \boldsymbol{n}\\
\text{subject to\quad} & A^T \boldsymbol{n}\leq \boldsymbol{b}
\quad \text{and} \quad \boldsymbol{n} \geq 0,
\end{split}
\label{eq:IP_definition}
\end{equation}
where $\boldsymbol{n}$ is a column vector of $N$  integer variables, $c$ is the matrix of coefficients describing the objective function, and the matrix $A$ and column vector
$\boldsymbol{b}$ specify a system of $\ell$ inequality constraints on the variables. As mentioned
above, we use the boson number observable for each of the involved bosonic modes to represent the corresponding
entries of $\boldsymbol{n}$. Hence, the condition $\boldsymbol{n} \geq 0$ and the integrality of these
variables are automatically satisfied.

Each inequality $a_j^T \boldsymbol{n} \leq b_j$, for $1 \leq j \leq \ell$, can be
rewritten as an equality constraint $a_j^T \boldsymbol{n} - b_j + \eta_j = 0$ via an
auxiliary variable $\eta_j \geq 0$. This requires the addition of $\ell$ ancillary
qumodes. Assuming $A$ and $\boldsymbol{b}$ have fractional entries, we can multiply them
by their lowest common denominator to turn all of their entries into integers. We then
use the number operator of the $j$-th ancillary qumode to represent $\eta_j$.
Alternatively, $\eta_j$ can be considered a continuous variable represented by
the non-negative-valued observable $x_j^2$ of the $j$-th qumode. This, however,
turns \mbox{problem \eqref{eq:IP_definition}} into an MIP
similar to those in \cref{sec:MIP}.

To solve problem \eqref{eq:IP_definition}, we seek the ground state of the
Hamiltonian
\begin{equation}
\hat H_\text{P}=
- \sum_{i=1}^{N}c_{i}\hat n_i
+ \lambda \sum_{j=1}^{\ell}
\left[\left(\sum_{i=1}^N A_{ij}\hat n_i\right)
+ \hat \eta_j- b_j\right]^{2},\label{eq:IP_H_T}
\end{equation}
that is, by replacing the vectors $\boldsymbol{n}$ and $\boldsymbol{\eta}$ with vectors of the photon number operators $\boldsymbol{\hat n}$ and $\boldsymbol{\hat \eta}$. Here, $\lambda>0$ is a penalty coefficient tuned to suppress
the infeasible subspace of the Hilbert space. We note that simulating the
problem \mbox{Hamiltonian \eqref{eq:IP_H_T}} requires the implementation of unitary
evolutions of the form $\exp(i\alpha \hat n_i \hat n_j)$, $\exp
(i\alpha \hat n_i^2)$, and $\exp(i\alpha \hat n_i)$, which can all be achieved
using cross-Kerr, Kerr, and rotation operations, respectively.

As a first example in our study of ILP problems, we consider the simple problem of solving the equation
\mbox{$n_1 + n_2 = 5$} over the integer domain. This simple equation can be re-written
in the canonical form \eqref{eq:IP_definition} by setting
$$A= \begin{pmatrix} 1 & -1\\ 1 & -1 \end{pmatrix},\;
\boldsymbol{b}= \begin{pmatrix} 5 \\ -5 \end{pmatrix},\;\text{ and }\;
c= \begin{pmatrix} 0 \\ 0 \end{pmatrix},
$$
that is, by imposing two inequalities, $n_1 + n_2 \leq 5$ and
$-n_1 -n_2 \leq -5$, to determine the feasible domain.
As such a problem does not effectively minimize an objective function but
  requires the finding of integer solutions only to a set of constraints, it is known
as a \emph{feasibility problem}. For this reason, we avoid reformulating the problem in the canonical form (as it would require employing four
qumodes) and instead find the ground states of
\mbox{$\hat H_\text{P}= (\hat n_1 + \hat n_2 - 5)^2$.} 

\Cref{fig:IP_adiabatic_final_states} shows the probability distribution of the results of measuring the final state in the two-mode Fock basis
prepared using AQC. The probability amplitudes of the quantum states evolved over time via AQC and through its discretized form are depicted in~\cref{fig:IP_adiabatic_evolution,fig:IP_QAOA_evolution}, respectively. The two methods yield similar results, with the most-likely measurement outcomes corresponding to the six degenerate Fock states $\ket{0,5}, \ket{1,4}, \ket{2,3}, \ket{3,2}, \ket{4,1}$, and $\ket{5,0}$ representing the solutions of the ILP problem. We observe that the ground-state amplitudes are amplified by the adiabatic
scheme while those of the other Fock states are suppressed. We tune $p_{0i}$ in the mixing Hamiltonian~\eqref{eq:Mixing} to achieve fair sampling (see \cref{sec:fair_sampling}).

We now consider a more interesting ILP problem in the canonical form. In the
\emph{knapsack problem}, a set of items and a knapsack are considered, where each item
$i$ has a value $v_i$ and a weight $w_i$ associated to it, and the
knapsack has a total capacity $W$. The goal is to insert a number of items into
the knapsack such that their selection maximizes the sum of the values of items
in the knapsack while not exceeding its capacity. In typical logistics
scenarios, however, each item is of a \mbox{\emph{type} $i$} and we are allowed to choose
any number of items of that type. This is called the \emph{unbounded} knapsack problem
(UKP), which can be formulated in the canonical form \eqref{eq:IP_definition} as follows:
\begin{equation}
\begin{split}
\text{minimize}\quad & -\sum_{i=1}^N v_i n_i\\
\text{subject to\quad} &
\sum_{i=1}^N w_i n_i \leq W,\quad n_i \geq 0 \text{ for all }i.
\end{split}
\label{eq:BKP_definition}
\end{equation}
The integer variables $n_i$ determine how many of each item type, if any,
can be included in the knapsack.

\begin{figure}[t]
\includegraphics[scale=0.24]
{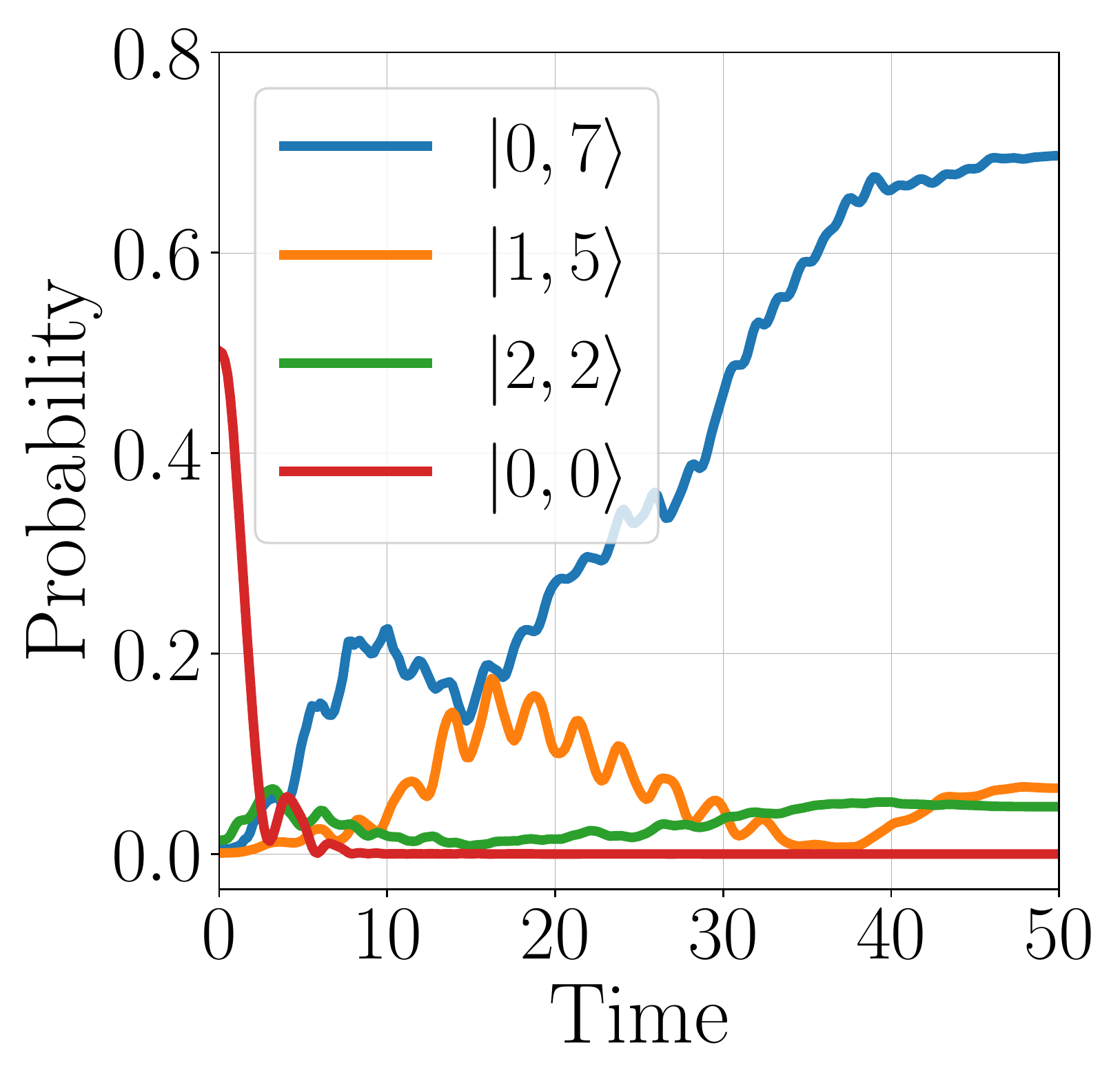}
\caption{State evolution
during the execution of  continuous AQC  to solve the unbounded knapsack problem instance~\eqref{eq:BKP_toy} using the total evolution time $T= 50$. The initial states are the squeezed coherent states
$\ket{\mathbf{\boldsymbol{\alpha}}, -{\boldsymbol{r}}}$, where
$\boldsymbol{\alpha}=(\alpha_1,\alpha_2,\alpha_3)$, with $\alpha_1 = \alpha_2= \alpha_3= \frac{0.25i}{\sqrt{2}}$ 
and $r =0.8$. Note that the third state corresponds to the ancillary qumode, which is traced over in calculating the probabilities at the end of AQC. The penalty coefficient used  is $\lambda = 4$.
\label{fig:evolutions_BKP}}
\end{figure}

\Cref{fig:evolutions_BKP} shows the time progress of AQC for solving the UKP instance
\begin{equation}
\begin{split}\text{minimize\quad} & -n_1 - 2n_2\\
\text{\text{subject to\quad}} & 4n_1+1.5n_2\le11,\quad n_1, n_2 \ge 0,
\end{split}
\label{eq:BKP_toy}
\end{equation}
which has a unique optimal solution $(n_1, n_2) = (0, 7)$. We observe that the state $\ket{\psi}=\ket{0,7}$ is amplified by AQC while all
other Fock states are suppressed.

\section{Nonlinear Integer Programming}
\label{sec:NLIP}

In NLIP problems, both the objective functions and the
constraints can be nonlinear functions of the integer variables. Such problems
can be converted into ILP problem instances; however, this is at the cost of requiring
additional qumodes. In this section, we show that NLIP problems can be solved
directly on a bosonic quantum computer.

\begin{figure}[b]
\subfloat[\label{fig:MaxClique_Example}]{
\begin{tikzpicture}
[scale= 0.5, auto= right,
every node/.style= {circle, draw=black, line width=0.1mm, fill= red!40}]
\node (n1) at (0,3)  {1};
\node (n2) at (1,0)  {2};
\node (n3) at (4,-1)  {3};
\node (n4) at (5,2)  {4};
\node (n5) at (3,4)  {5};
\foreach \from/\to in {n1/n2,n1/n3,n1/n4,n2/n4,n2/n5,n3/n4}
\draw [line width=0.25mm] (\from) -- (\to);
\end{tikzpicture}}
\hspace{5mm}
\subfloat[\label{fig:evolution_MC-QUBO_adiabatic}]
{\includegraphics[scale=0.24]
{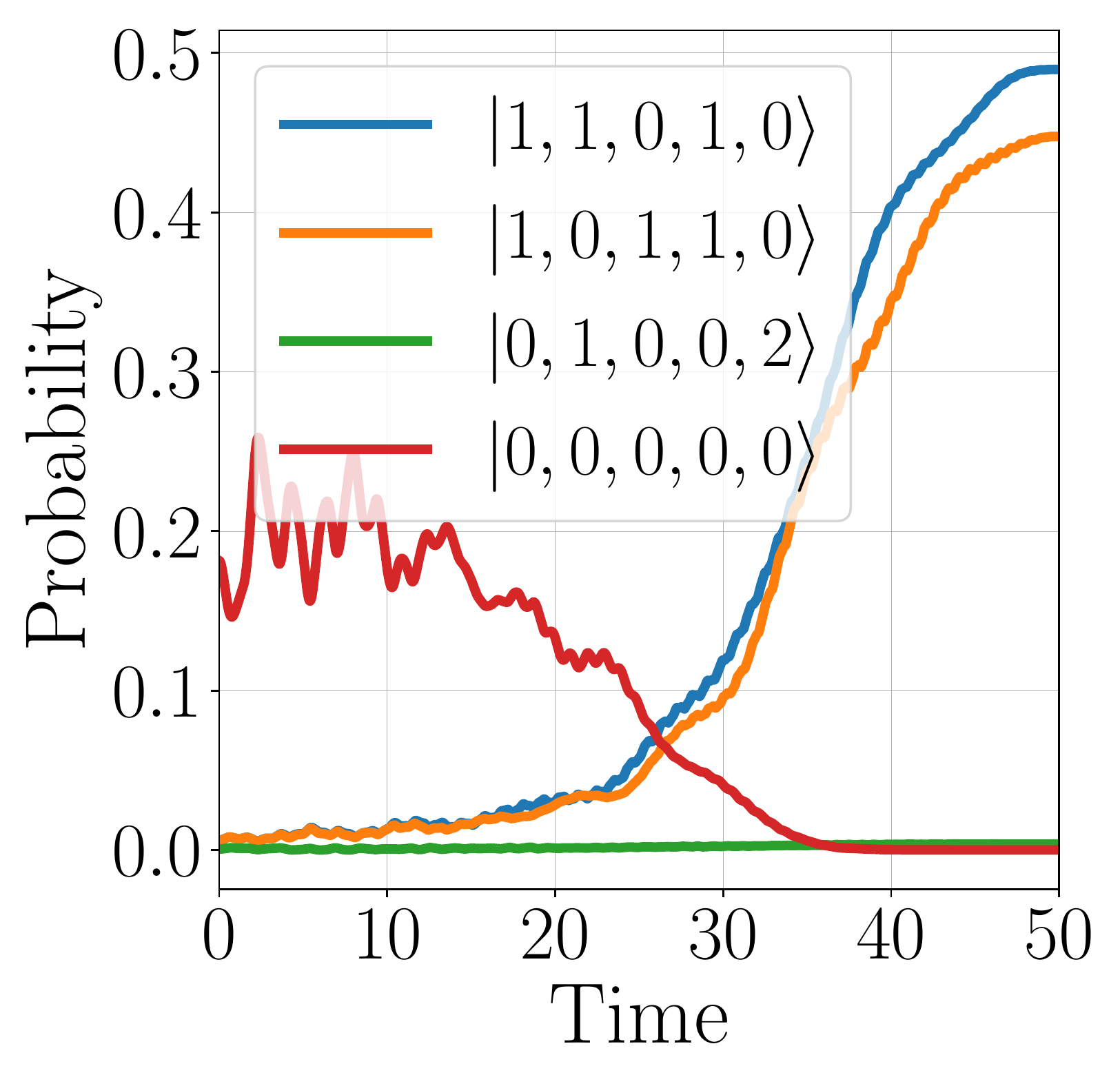}}

\caption{
(a) Example graph with five vertices, used as a \textsc{MaxClique} problem instance in our analysis. The graph contains two maximum
cliques that are the subgraphs determined by the vertices $\{1,2,4\}$ and
$\{1,3,4\}$.
(b) State evolution
during AQC to solve the \textsc{MaxClique} problem instance. We use the problem Hamiltonian
\eqref{eq:MaxClique-Binary} with $\lambda = 1$ and $\mu = 6$,
and simulate the evolution of five qumodes all initialized as the  squeezed coherent states
$\ket{\mathbf{\boldsymbol{\alpha}}, -{\boldsymbol{r}}}$, where
$\boldsymbol{\alpha}=(\alpha_1,\alpha_2,\ldots,\alpha_5)$, with  $\alpha_j=0.55i/\sqrt{2}$ for all $j$ and $r=0.5$, for a total
evolution time of $T=50$. The degenerate ground states are
$\ket{1, 1, 0, 1, 0}$ and $\ket{1, 0, 1, 1 ,0}$. The next largest probability
amplitude (green curve) observed has two photons in the fifth qumode.
However, this state is not a ground state of Hamiltonian~\eqref{eq:MaxClique-Binary}. The
probability of the vacuum state $\ket{0,0,0,0,0}$ is shown
using a red curve.
\label{fig:QUBO-MaxClique}}
\end{figure}
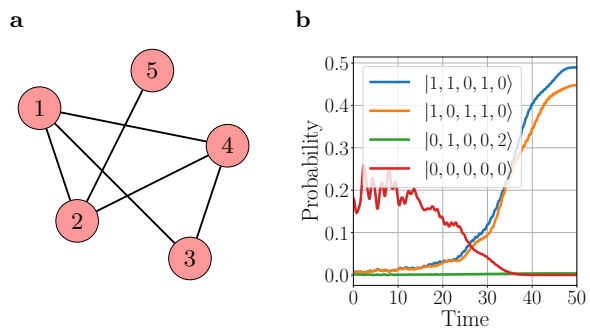

We use the \textsc{MaxClique} problem \cite{Bomze1999} as our working example.
In this section, we consider a binary NLIP formulation of this problem, and in \cref
{sec:Non-convex} we solve an integer NLIP formulation as well as a non-convex continuous formulation. In the \textsc{MaxClique} problem, a graph $G = (V, E)$ is given, where $V$ is the set of vertices of $G$ and $E$ is the set of edges. Each
edge $e \in E$ is a pair of vertices $\{i, j\} \subseteq V$. The set of edges can be represented by a symmetric adjacency matrix $A$, wherein $A_{ij}=1$ if $\{i, j\}$ is in $E$, and
$A_{ij}=0$ otherwise. The goal of the \mbox{\textsc{MaxClique}} problem is to find the largest {\em complete} subgraph of $G$, that is, to find the largest subset $S \subseteq V$ such that all pairs of vertices
in $S$ are connected by an edge. In what follows, we  also use $S$ to denote
the subgraph it determines (see~\cref{fig:MaxClique_Example} for an example).

Let $n_i \in \{0, 1\}$ represent the selection of vertex $i$ in the maximum
clique. Then, the following NLIP formulation can be used to solve the \textsc{MaxClique} problem:
\begin{equation}
\begin{split}\text{minimize}\quad & -\sum_{i}n_i\\
\text{subject to}\quad & \boldsymbol{n}^T (\boldsymbol{1} - A) \boldsymbol{n} = 0,
\\
& n_i \in \{0, 1\} \text{ for all }i.
\end{split}
\label{eq:MaxClique-Binary}
\end{equation}
Here, $\boldsymbol{1}$ is the $N\times N$ matrix with every element  equal to $1$. We thus write the following problem
Hamiltonian in terms of the number operators of the qumodes:
\begin{equation}
\hat H_\text{P}= - \sum_i \hat n_i
+ \lambda \sum_{i,j} (1- A_{ij}) \hat n_i \hat n_j
+ \mu \sum_i \hat n_i(\hat n_i- 1)\,.
\label{eq:target_H_QUBO}
\end{equation}
The first term corresponds to the objective function of problem~\eqref{eq:MaxClique-Binary} and the second term penalizes any violation of the
adjacency constraint of problem~\eqref{eq:MaxClique-Binary}. The last term restricts
our search to the subspace of the Fock space corresponding to the eigenvalues $0$ or
$1$ for the number operators of the qumodes. It is easy to see that any values of 
$\lambda > \frac{1}2$ and $\mu > 1$ suffice.

We apply AQC to the \textsc{MaxClique} problem for the 
graph shown in  \cref{fig:MaxClique_Example}. In
\cref{fig:evolution_MC-QUBO_adiabatic}, we plot the three largest probabilities at the end of the adiabatic evolution. The two degenerate states representing the maximum
cliques correspond to the vertices $\{1,2,4\}$ and $\{1,3,4\}$. Thus, in \cref
{fig:evolution_MC-QUBO_adiabatic}, the probability amplitudes of the five-mode
Fock states $\ket{1,1,0,1,0}$ and $\ket{1,0,1,1,0}$ converge to large values.

\section{Non-convex global optimization}
\label{sec:Non-convex}

Non-convex continuous functions are difficult to optimize classically even
when they contain no discrete variables~\cite{burer2012non}. For instance, Motzkin and Straus~\cite{motzkin1965maxima} provide a continuous formulation of the NP-hard \textsc{MaxClique} problem. As in the previous section, we let $A$ denote the adjacency matrix
of a graph $G$. Then, according to the \mbox{Motzkin--Straus} theorem, the optimal solutions of the
continuous quadratic programming problem,
\begin{equation}
\begin{split}
\text{minimize\ensuremath{\quad}} & -x^T A x\\
\text{subject to\ensuremath{\quad}} & \sum_i x_i= 1,\quad x_i \geq 0 \text{ for all } i,
\end{split}
\label{eq:motzkin_continuous}
\end{equation}
have nonzero entries $x_i = \frac{1}{\xi}$ for $i \in S$, where $S$ is a
maximum clique of $G$,  and $x_i= 0$ for all $i \not\in S$. Here, $\xi = |S|$ is the size of the one or possibly many maximum cliques.

We use the position observable of the qumodes to represent continuous variables,
and to impose the non-negativity of each $x_i$ we substitute $\hat x_i$ with
$\hat x_i^2$. The resulting problem Hamiltonian is
\begin{equation}
\hat H_\text{P}= -\sum_{i,j} A_{ij} \hat x_i^2 \hat x_j^2
+\lambda \left(\sum_i \hat x_i^2 -1\right)^2.
\label{eq:target_MS_continuous}
\end{equation}
As all entries in the optimal solutions of problem~\eqref{eq:motzkin_continuous} are
$0$ or $1/\xi$, we may substitute each $x_i$ with $n_i/\xi$ for an
integer variable $n_i$ and instead solve the problem
\begin{equation}
\begin{split}
\text{minimize}\quad & -n^T A n\\
\text{subject to\ensuremath{\quad}} & \sum_i n_i = \sigma,
\quad n_i \in \mathbb Z_{\geq 0} \text{ for all }i,
\end{split}
\label{eq:motzkin_integer}
\end{equation}
using $\sigma = \xi$. However, $\xi$ is
not a priori known, but we may perform a binary search for its value using $\mathcal{O}(\log (|V|))$ choices of $\sigma \in \{1, \ldots, |V|\}$, where $|V|$ is the number of vertices of the graph. In
addition, we can infer the maximum cliques correctly even when $\sigma \geq \xi$
(see \cref{sec:MS-hyperparam} for more detail). Using an  approach similar to that of \cref{sec:NLIP}, the
resulting problem Hamiltonian for this NLIP can be written as
\begin{equation}
\hat H_\text{P}= -\sum_{i,j} A_{ij} \hat n_i \hat n_j
+ \lambda \left(\sum_i \hat n_i - \sigma \right)^2.
\label{eq:target_MS_discrete}
\end{equation}
We note that the Hamiltonians \eqref{eq:target_H_QUBO}, \eqref{eq:target_MS_continuous}, and \eqref{eq:target_MS_discrete} are all quartic
in $a$ and $a^{\dagger}$ because $\hat n$ and $\hat x^2$ are both quadratic in
$a$ and $a^{\dagger}$.

\begin{figure}[!b]
\subfloat[\label{fig:MS-continuous}]
{\includegraphics[scale=0.24]
{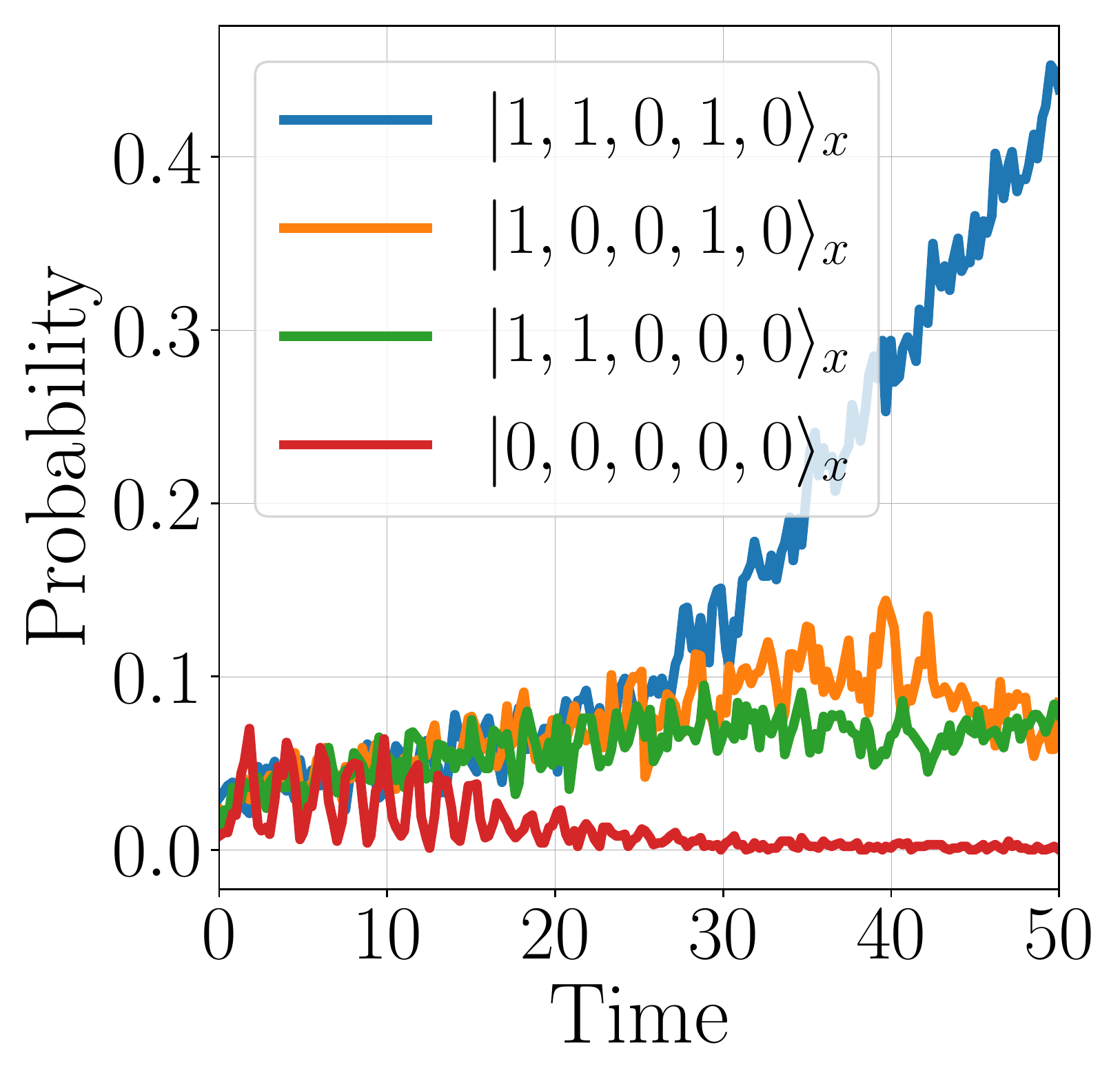}}
\subfloat[\label{fig:MS-discrete}]
{\includegraphics[scale=0.24]
{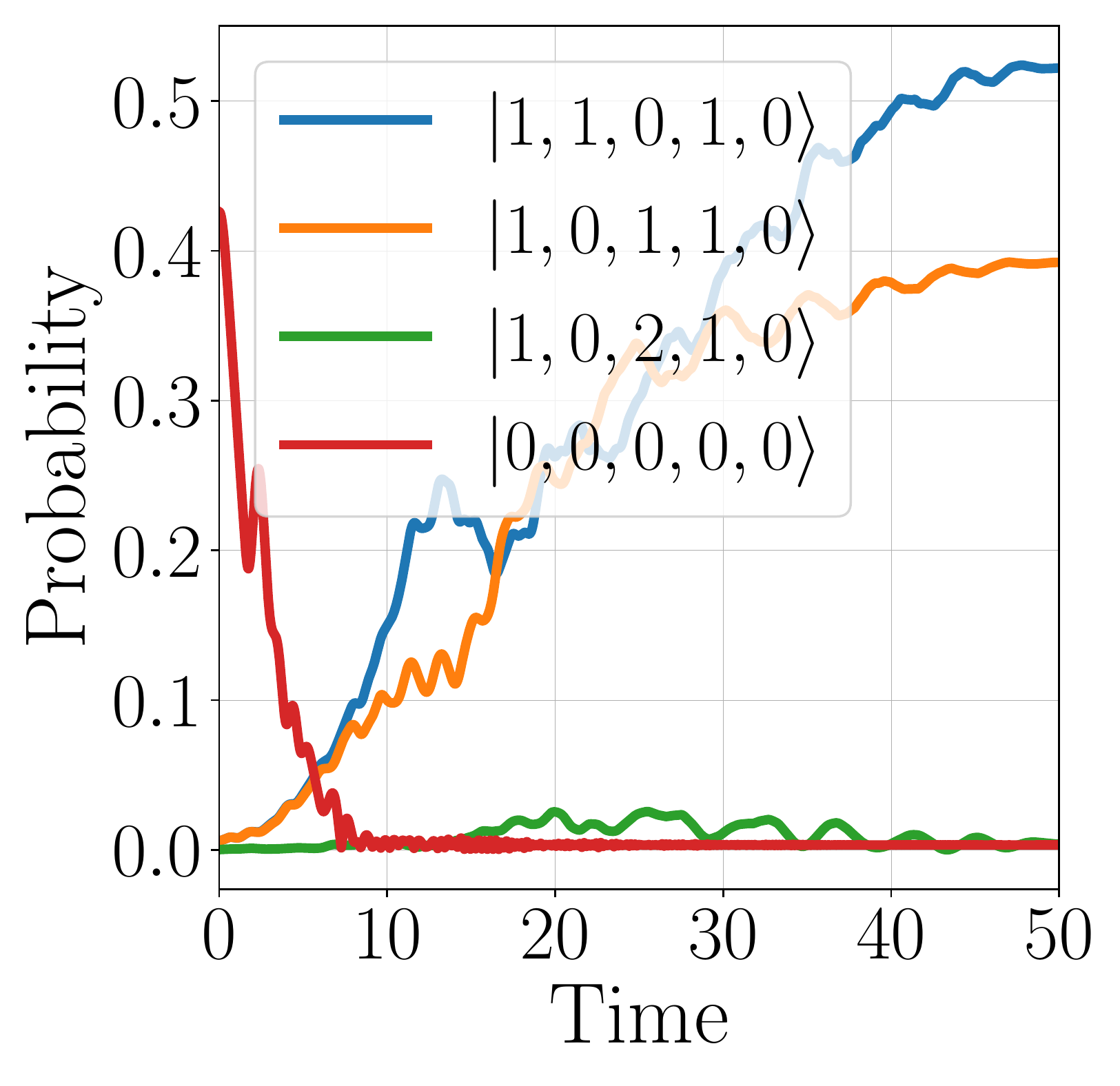}}
\caption{Probability evolution in solving the \text{MaxClique} problem using the Motzkin--Straus framework, for the graph shown in
\cref{fig:MaxClique_Example}. (a) Solving the non-convex continuous quadratic problem \eqref{eq:motzkin_continuous} using the problem Hamiltonian
\eqref{eq:target_MS_continuous}. (b) Solving the non-convex integer quadratic problem~\eqref{eq:motzkin_integer} using the problem Hamiltonian~\eqref{eq:target_MS_discrete}. In both (a) and (b), the five qumodes
are initialized as the squeezed coherent states $\ket{\mathbf{\boldsymbol{\alpha}},-\boldsymbol{r}}$, where
$\boldsymbol{\alpha}=(\alpha_1,\alpha_2,\ldots,\alpha_5)$, with
$\alpha_j=0.55i/\sqrt{2}$ for all $j$ and $r=0.5$, and evolved using AQC for a total evolution time of $T= 50$. In
both formulations, $p_{0}=0.55\hbar$ in the mixing
Hamiltonian. The penalty coefficient is $\lambda = 2$ for the continuous
formulation and $\lambda = 6$ for the integer formulation.
\label{fig:Motzkin-Straus}}
\end{figure}

In \cref{fig:MS-continuous}, we show the progression of AQC using the problem
Hamiltonian \eqref{eq:target_MS_continuous} for solving the \textsc{MaxClique} problem for the graph in \cref
{fig:MaxClique_Example}. The solutions are found by performing a homodyne
measurement of the $\hat{x}$ observable for each mode, respectively 
(i.e., by projection on the $\ket{x}$
eigenstates). Here, $\ket 1_x$ denotes a state with a measured value that is
greater than or equal to $1/\sqrt{2|V|}$ for the eigenvalue of $\hat x$, while $\ket 0_x$ denotes a state with 
 a measured value less than $1/\sqrt{2|V|}$. The probabilities are found by performing measurements 1000 times. The five qumodes are measured 
 consecutively with conditional homodyne measurements, and a histogram 
 of the states $\ket{n_1, n_2, n_3, n_4, n_5}_x$ is generated, where $n_i \in \{0, 1\}$ 
 according to the above definition. The probabilities are 
 found from this histogram.

In \cref{fig:MS-discrete}, we show the evolution of the probabilities of both the dominant
quantum states (in this case those that have the highest final probability amplitudes) and the vacuum state, after performing AQC
using the NLIP representation of the problem Hamiltonian~\eqref{eq:target_MS_discrete}. We observe that the
continuous evolution shown in~\cref{fig:MS-continuous} is less performant (with a total
success probability of observing a maximum clique of about $55\%$) than the probabilities shown in~\cref{fig:evolution_MC-QUBO_adiabatic} and~\cref{fig:MS-discrete} for the integer formulations of the \textsc{MaxClique} problem (both
with success probabilities greater than $90\%$). One reason for this is the penalty relaxation in the Hamiltonian~\eqref{eq:target_MS_continuous} is not
equivalent to the original problem~\eqref{eq:motzkin_continuous} and instead
provides a lower bound for the optimal value of problem~\eqref{eq:motzkin_continuous}. Another reason is that, in our numerical simulations, the Fock space has been
truncated to five dimensions due to limitations in classical computational resources. These two facts point to there being a
broader probability distribution in the $x$ direction of the phase space, which can cause a
greater variance in the $\hat{x}$ measurements.

We also observe a greater number of oscillations over time in the ground-state amplitudes
for problem~\eqref{eq:motzkin_integer} compared to the binary formulation~\eqref{eq:MaxClique-Binary}. This can be understood by the fact that Hamiltonian~\eqref{eq:target_MS_discrete} has larger eigenvalues than  Hamiltonian~\eqref{eq:target_H_QUBO}; the evolution via Hamiltonian~\eqref{eq:target_MS_discrete} is
populated by higher-number Fock states, whereas in Hamiltonian~\eqref
{eq:target_H_QUBO} only the values $0$ or $1$ are allowed for the boson number observable of each qumode. The amplitude oscillations can therefore be reduced by scaling down the
problem Hamiltonian by a factor that is less than one while leaving the mixing Hamiltonian unaffected. Experimentally, performing such a normalization can
be helpful in situations where the parameters of the quantum optical gates
must remain within specific operating ranges.
Finally, we note that the difference between the probabilities of observing
the different degenerate ground states in~\cref{fig:evolution_MC-QUBO_adiabatic} and \cref{fig:MS-discrete} can be
reduced by tuning the parameters of the initial states and the mixing
Hamiltonian in order to achieve more ``fair'' samples. See \cref{sec:fair_sampling} for more detail.

\section{Mixed-Integer Programming}
\label{sec:MIP}

\begin{figure*}[t]
\subfloat[\label{fig:sparse-AQC}]{\includegraphics[scale=0.205]
{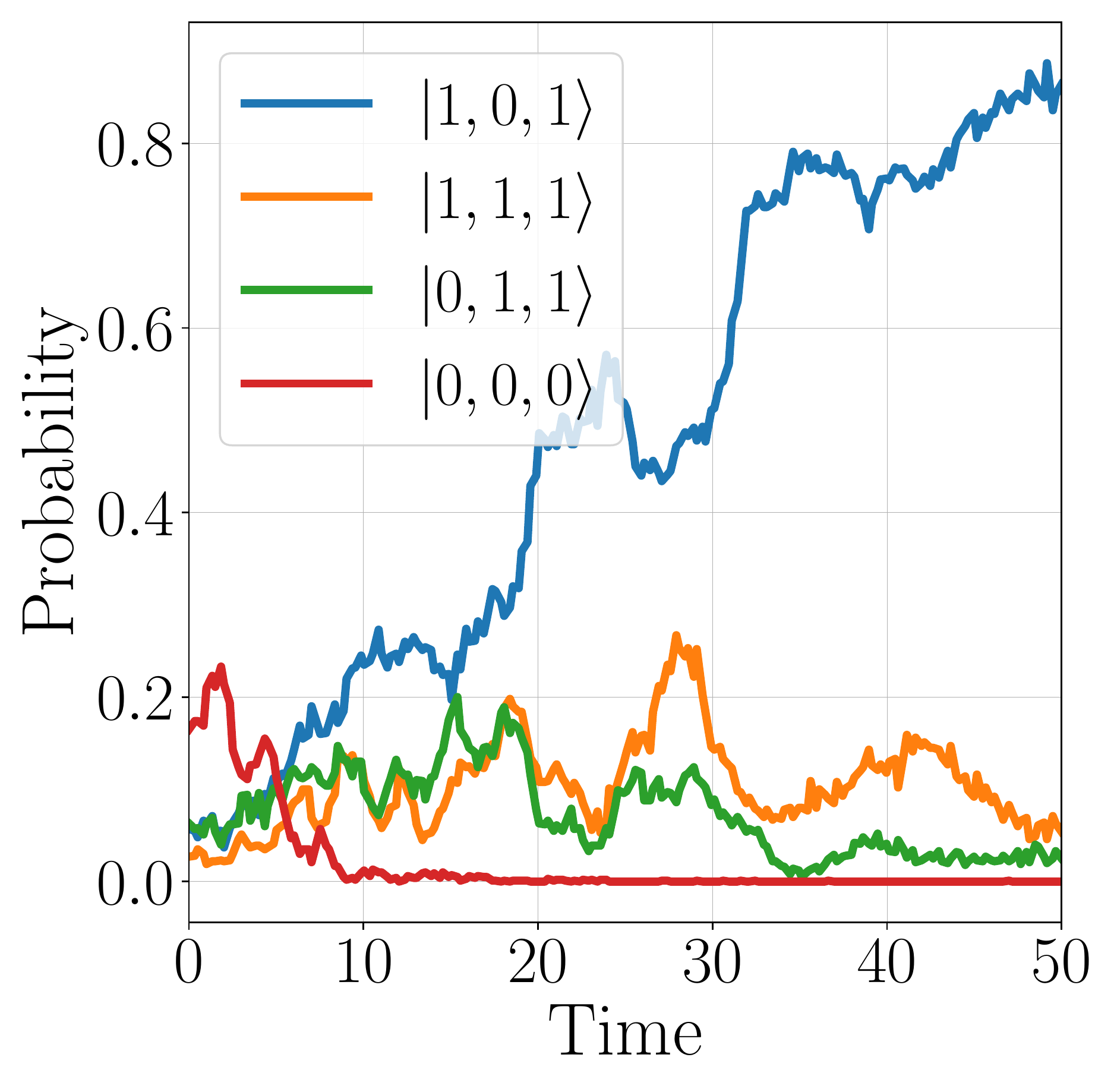}}
\hspace{1mm}
\subfloat[\label{fig:sparse-AQC-prob-T0}]{\includegraphics[scale=0.25]
{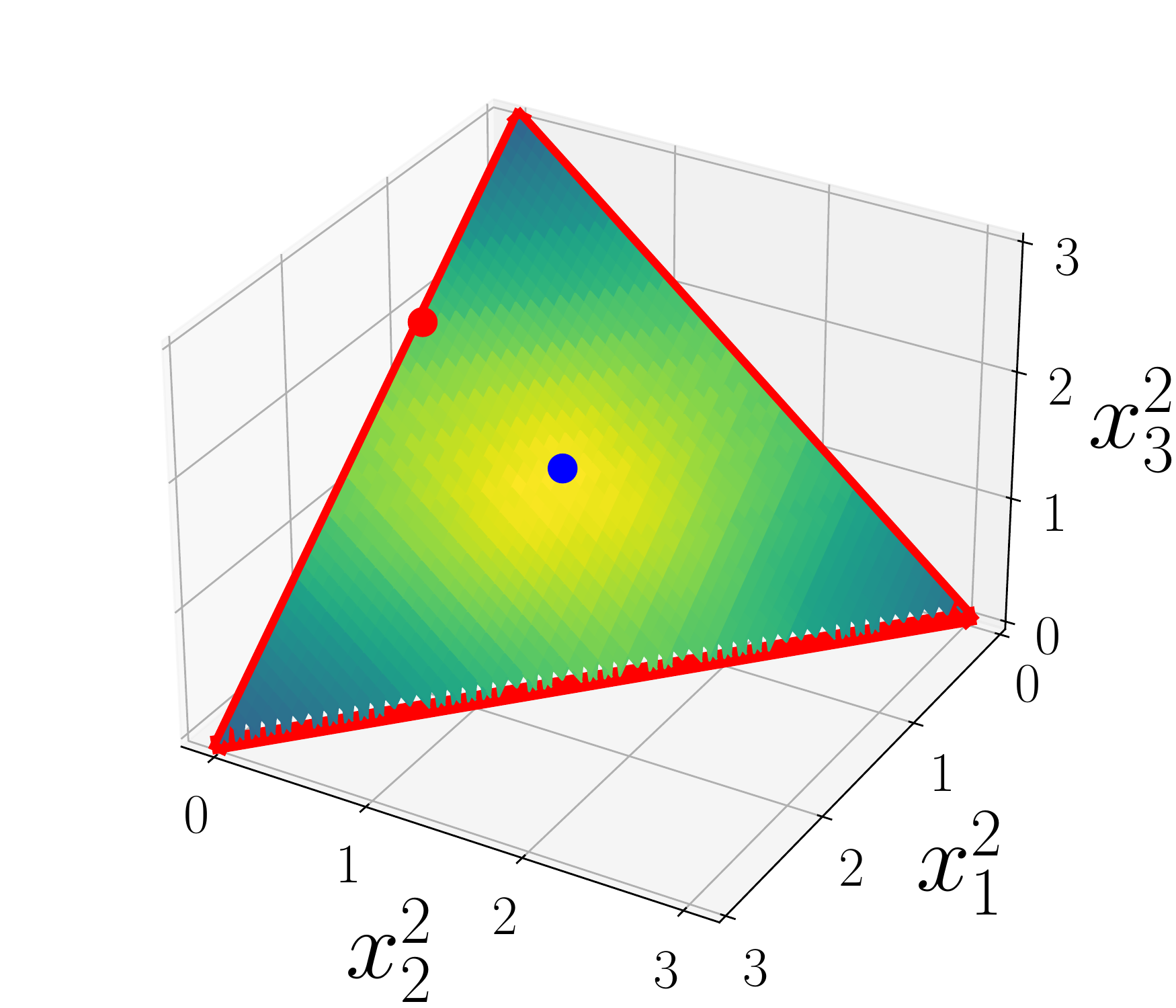}}
\hspace{1mm}
\subfloat[\label{fig:sparse-AQC-prob-T0.5}]{\includegraphics[scale=0.25]
{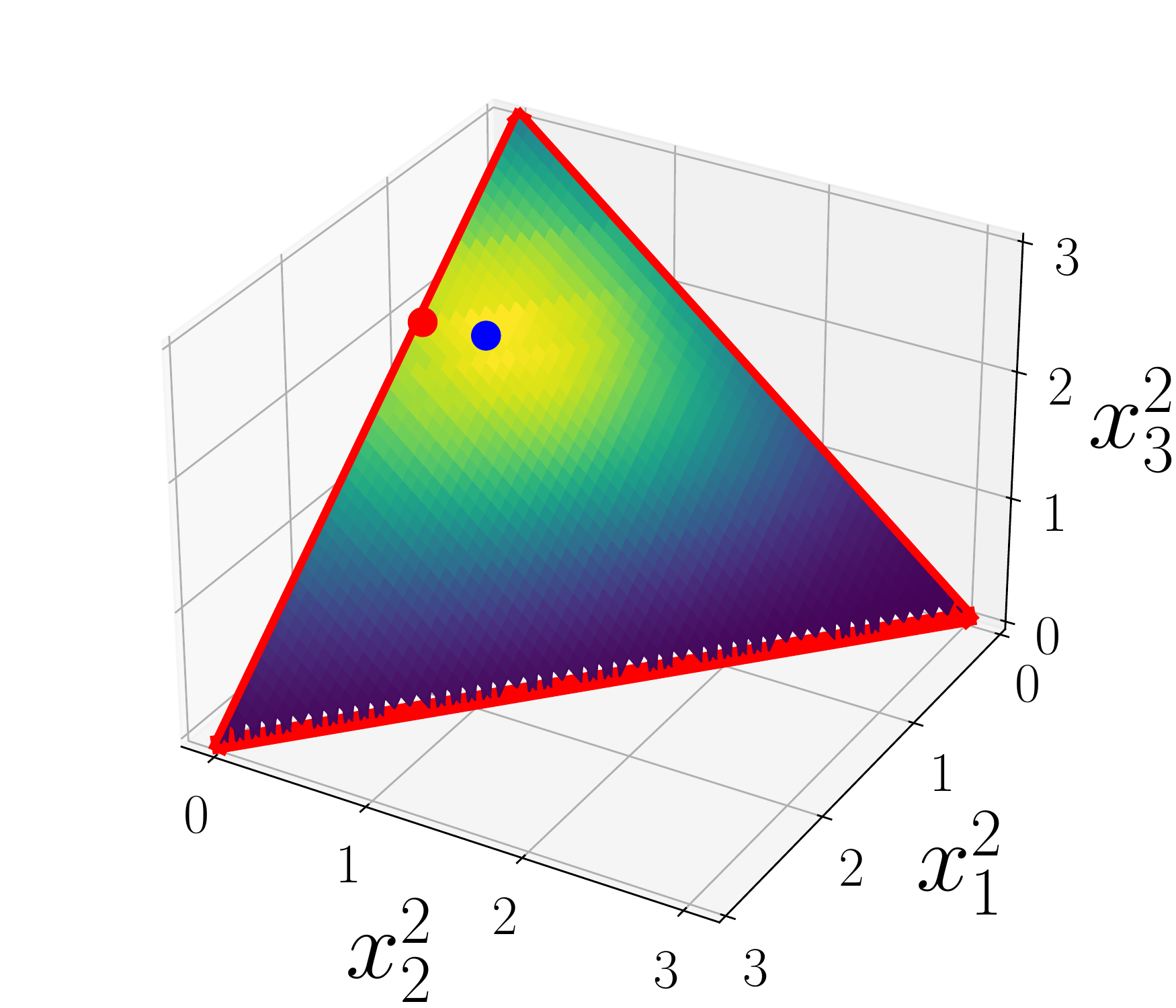}}
\hspace{1mm}
\subfloat[\label{fig:sparse-AQC-prob-T1}]{\includegraphics[scale=0.25]
{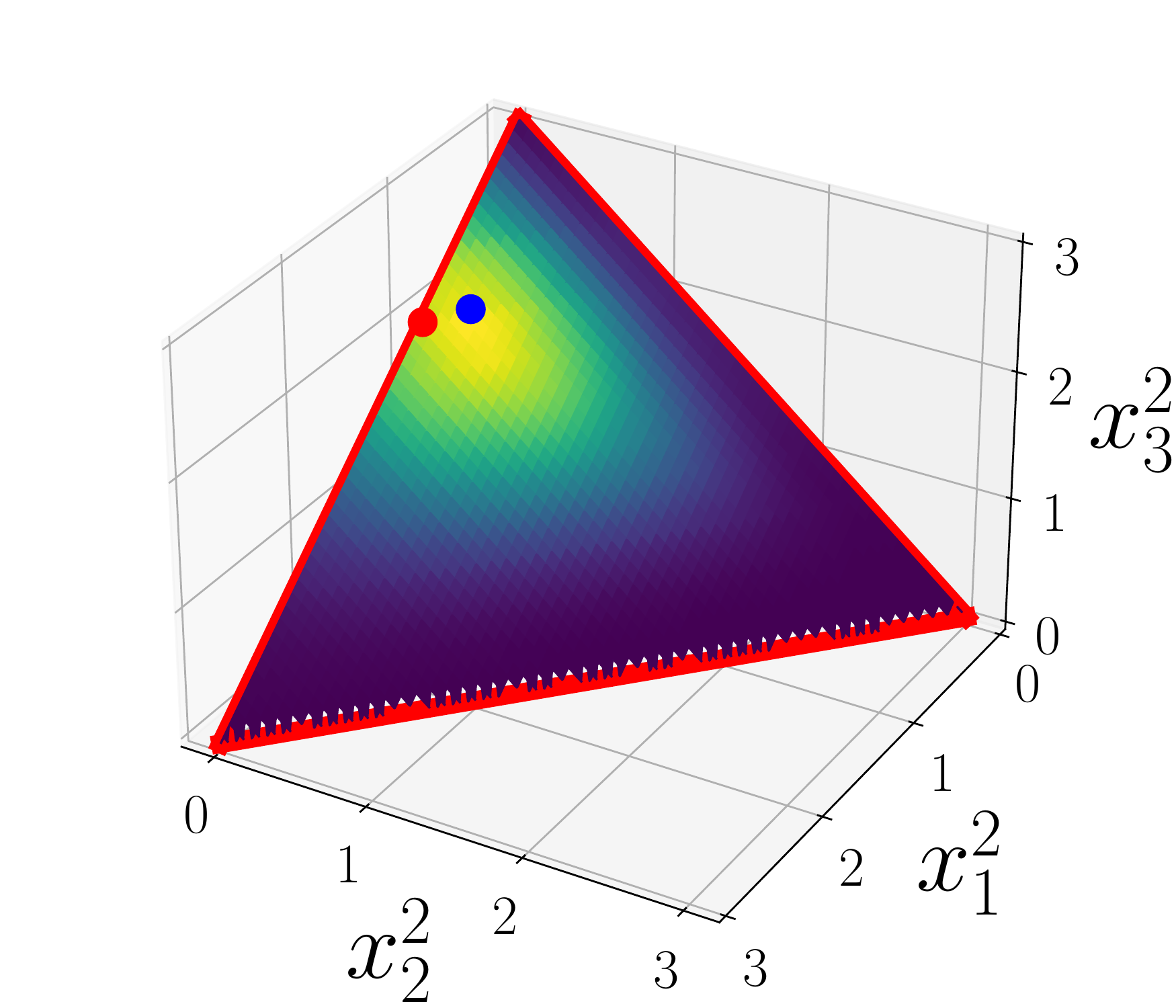}}

\caption{Progression of AQC in finding the ground states of the problem Hamiltonian~\eqref{eq:sparse-target} representing the MIP problem~\eqref{eq:sparse_MIP}. We use $(\lambda_1, \lambda_2, \lambda_3) = (1.2, 0.3, 0.3)$
and a total evolution time of $T= 50$. All initial coherent states
have the squeezing parameter $r=-0.8$, and $p_{0}=0.55\hbar$, for all of the qumodes.
(a) Time evolution of the probabilities of the states $\ket{n_4, n_5, n_6}$ that have the highest
probabilities at the end of the adiabatic evolution, traced over the first three qumodes. The average measured values of the observables $\hat{x}_i^2$ for the first three qumodes, conditional upon the most probable quantum state $\ket{n_4, n_5, n_6} = \ket{1,0,1}$, at the end of AQC are $(\langle \hat{x}_1^2 \rangle, \langle \hat{x}_2^2 \rangle, \langle \hat{x}_3^2 \rangle) \simeq (0.7, 0.4, 1.6)$. The values
of $\langle{\hat x_i^2}\rangle$ for these qumodes are calculated by averaging 1000 homodyne measurements.
(b)--(d) The time evolution of the probability distribution of the
first three qumodes representing the continuous variables of the sparse optimization problem. The panels show the instantaneous probability distribution at
(b) $t=0$, (c) $t=T/2$, and (d) $t=T$, on the simplex. The discrepancy between the optimal solution (shown using a red dot) and
the maximum of the probability distribution (blue dot) at $t=T$ is due to the truncation of the Fock space. Note that the variance of the distribution decreases throughout the adiabatic evolution.
\label{fig:time-evolution_sparse_simplex}}
\end{figure*}

Finally, we demonstrate the application of our method in solving MIP problems (i.e., problems that consist of both integer and
continuous variables). We use sparse optimization as our example.

Consider the problem
\begin{equation}
\begin{split}
\text{minimize}\quad & f(x)=\sum_{i=1}^{3} (x_i-\mu_i)^{2}\\
\text{subject to\ensuremath{\quad}}
& \sum_{i=1}^3 x_i= 3,\quad \|x\|_0 \leq 2,\\
& x_i \geq 0 \text{ for all } i.
\end{split}
\label{eq:sparse_opt}
\end{equation}
Here, $\mu_i$ are constants. The objective function is a quadratic
(convex) function of three continuous variables defined on the (also convex)
positive two-dimensional simplex
$S= \{(x_1, x_2, x_3): \sum_{i=1}^3 x_i= 3\}$ in three dimensions. However, the
feasible domain for the problem is the 2-skeleton of $S$ consisting of those
points in $S$ that have at most two nonzero components,
$S \cap \{x: \|x\|_0 \leq 2\}$. As a result, the constraints impose a non-convex
feasible domain on the problem. Non-convex optimization problems as such are of
practical significance in regression, machine learning, and signal processing~\cite{tillmann2021cardinality}.

We reformulate problem~\eqref{eq:sparse_opt} as an MIP problem by adding the binary decision variables
$b_i$: 
\begin{equation}
\begin{split}
\text{minimize}\quad & f(x)=\sum_{i=1}^{3} (x_ib_i-\mu_i)^{2}\\
\text{subject to\ensuremath{\quad}}
& \sum_{i=1}^3 x_ib_i= 3,\quad \sum_{i=1}^3 b_i= 2,\\
& b_i \in \{0, 1\} \text{ and } x_i \geq 0 \text{ for all } i.
\end{split}
\label{eq:sparse_MIP}
\end{equation}
The assignement $b_i = 1$ indicates that the corresponding continuous variable $x_i$ is allowed to attain nonzero values. To solve this MIP problem using our method, we use six qumodes with the operators
$\hat x_i^2$, for $i= 1, 2, 3$, on the first three qumodes representing the continuous
variables, and the number operators $\hat n_i$, for $i = 4, 5, 6$,
representing the binary variables $b_{i-3}$. The MIP problem Hamiltonian
is given by
\begin{equation}
\begin{split}
\hat H_\text{P}= &
\sum_{i=1}^{3}(\hat x_i^2 \hat n_{i+3} - \mu_i)^2
+ \lambda_1 \left(\sum_{i=1}^3 \hat x_i^2 \hat n_{i+3} - 3\right)^2 \\ &
+ \lambda_2 \left(\sum_{i=1}^3 \hat n_{i+3} - 2\right)^{2}
+ \lambda_3 \sum_{i=1}^3 \hat n_{i+3}(\hat n_{i+3}-1).
\end{split}
\label{eq:sparse-target}
\end{equation}
Performing a homodyne measurement on the first three qumodes of the ground state
of this Hamiltonian returns the solutions $x_i$, while a photon number resolving measurement on the second three qumodes reveals the nonzero support
of the solution; that is, $x_i$ is taken to be zero if $n_{i+3}$ is zero.

\Cref{fig:time-evolution_sparse_simplex} shows the progression of AQC in finding the ground states of the problem Hamiltonian~\eqref{eq:sparse-target} representing the MIP problem~\eqref{eq:sparse_MIP}. \Cref{fig:sparse-AQC} shows the time evolution of the quantum states
$\ket{n_4, n_5, n_6}$ when traced over the first three qumodes. We use the constants $(\mu_1, \mu_2, \mu_3)= (1.0, 0.3, 2.0)$ in problem~\eqref{eq:sparse_opt} for our example problem instance, yielding the optimal solution
$(1.0, 0.0, 2.0)$. The quantum state $\ket{n_4, n_5, n_6}$ with the highest
probability at the end of the adiabatic evolution is $\ket{1,0,1}$, which implies that $x_2$ should be taken to equal zero. Together with the measured values of the position operators for the first three qumodes, we infer that the solution is
\mbox{$(x_1, x_2, x_3) \simeq (0.7, 0.0, 1.6)$} for the MIP problem instance. The discrepancy with the optimal
solution is due to the truncation of the Fock space to five dimensions. We
have also verified that increasing the Fock space truncation from four to five dimensions reduces this discrepancy. Finally,~\cref{fig:sparse-AQC-prob-T0} through~\cref{fig:sparse-AQC-prob-T1} show the time evolution
of the probability distribution of the $\hat x_i^2$ observables of the first three
qumodes on the simplex.

\section{Conclusion}

We have proposed a method for solving mixed-integer programming (MIP) problems using a set
of programmable bosonic quantum field modes. Our approach takes advantage of the fact that
the eigenspectrum of the bosonic number operators $\hat n= \hat a^\dagger \hat a$ consists of
non-negative integers; thus, they can naturally represent integer variables.
On the other hand, the quadrature operators $\hat{x}=\frac{1}{\sqrt{2}}\left(\hat{a} + \hat{a}^\dagger\right)$ and $\hat{p}= \frac{1}{i\sqrt{2}}\left(\hat{a} - \hat{a}^\dagger \right)$ have continuous
spectra and do not commute with each other or with the number operator.
Therefore, the MIP problem reduces to the problem of preparing the
ground state of a Hamiltonian written in terms of $\hat n$ and one of the quadrature
operators. For instance, an adiabatic ground-state preparation scheme can be
envisaged that uses a mixing Hamiltonian written in terms of the canonically conjugate
quadrature operator of the one used in the Hamiltonian specifying the problem.

We have discussed the experimental realization and numerical simulation of this scheme
on a continuous-variable, circuit model based quantum optical computer capable of
preparing coherent Gaussian states, and performing rotation gates, Kerr gates, and
cross-Kerr gates on them. At the end of the ground-state preparation scheme,
homodyne and photon number resolving  measurements were performed to return the
optimal solutions to an MIP problem. We further analyzed a
variety of linear and nonlinear MIP problems and provided
numerical demonstrations on small instances of \mbox{NP-hard} optimization problems,
specifically, the unbounded knapsack problem, the \textsc{MaxClique} problem, and
cardinality-constrained optimization.

This work demonstrates that bosonic continuous-variable quantum computation devices can provide the
natural processing capabilities required to solve MIP
problems efficiently. Our results, moreover, motivate the experimental
realization of a variety of non-Gaussian unitary gates on quantum optical
platforms. In addition, our work illuminates an interesting direction for future theoretical research in investigating the computational complexity of the scheme proposed herein in view of adiabatic theorems for Hamiltonians with continuous spectra (similar to~\cite{maamache2008adiabatic}).

\section*{Acknowledgement}
The authors thank our editor, Marko Bucyk, for his careful review and editing of the manuscript. The authors acknowledge the financial support received through the NSF’s CIM Expeditions award (CCF-1918549). P.~R.~acknowledges the financial support of Mike and Ophelia Lazaridis, and Innovation, Science and Economic Development Canada.
\clearpage
\onecolumngrid

\setcounter{section}{0}
\setcounter{equation}{0}
\setcounter{figure}{0}
\setcounter{table}{0}
\setcounter{page}{1}

\renewcommand{\theequation}{S\arabic{equation}}
\renewcommand{\thefigure}{S\arabic{figure}}
\renewcommand{\thetable}{S\arabic{table}}
\renewcommand{\thesection}{S\arabic{section}}
\renewcommand{\bibnumfmt}[1]{[#1]}
\renewcommand{\citenumfont}[1]{#1}

\begin{center}
\textbf{\large Supplementary Information:\\
Mixed-Integer Programming Using a Bosonic Quantum Computer}
\end{center}

\section{Adiabaticity}
\label{sec:adiabaticity}

\begin{figure*}[b]
\subfloat[]{\includegraphics[scale=0.3]{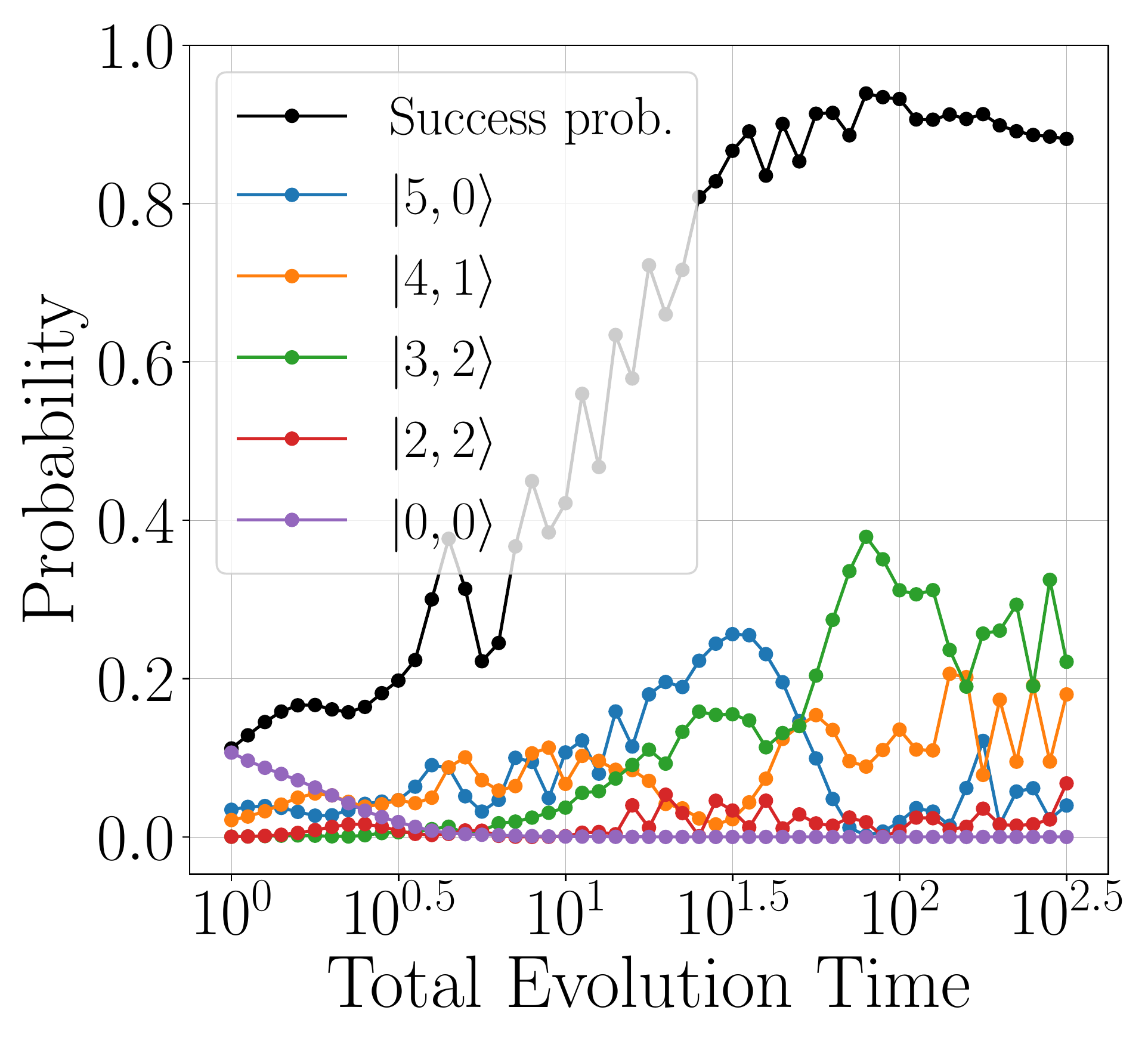}

}\subfloat[]{\includegraphics[scale=0.3]{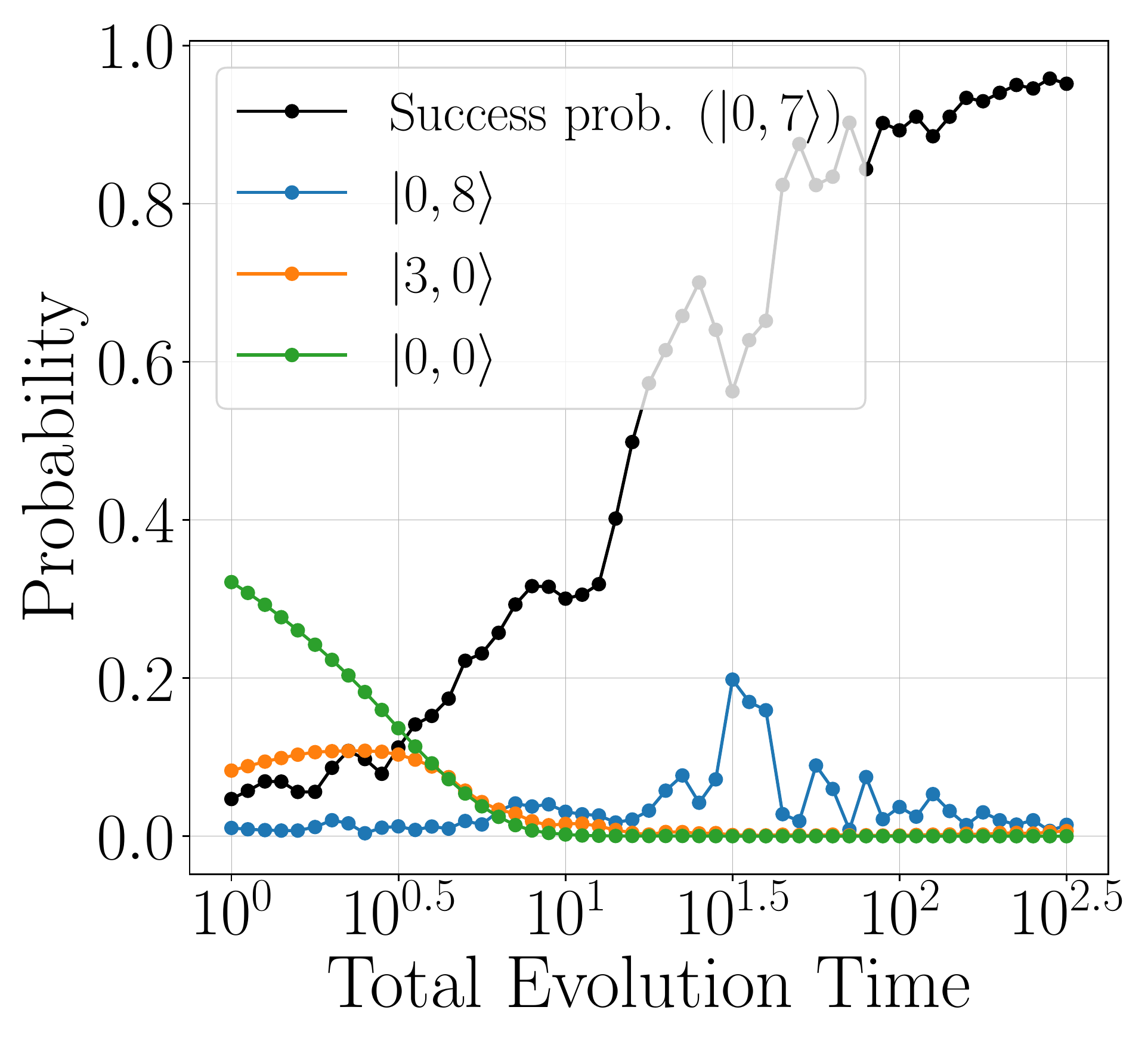}

}\subfloat[]{\includegraphics[scale=0.3]{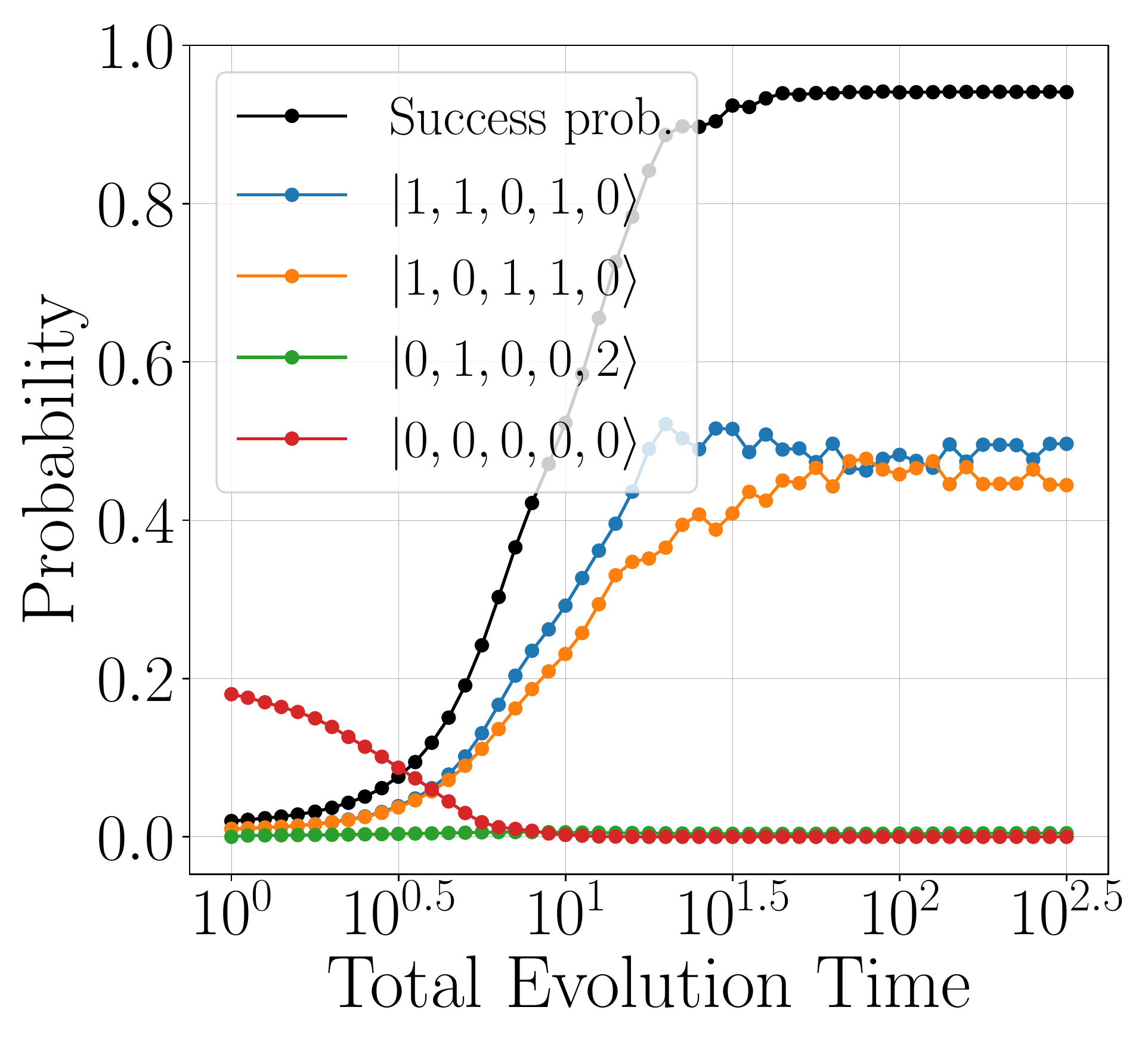}

}

\subfloat[]{\includegraphics[scale=0.3]{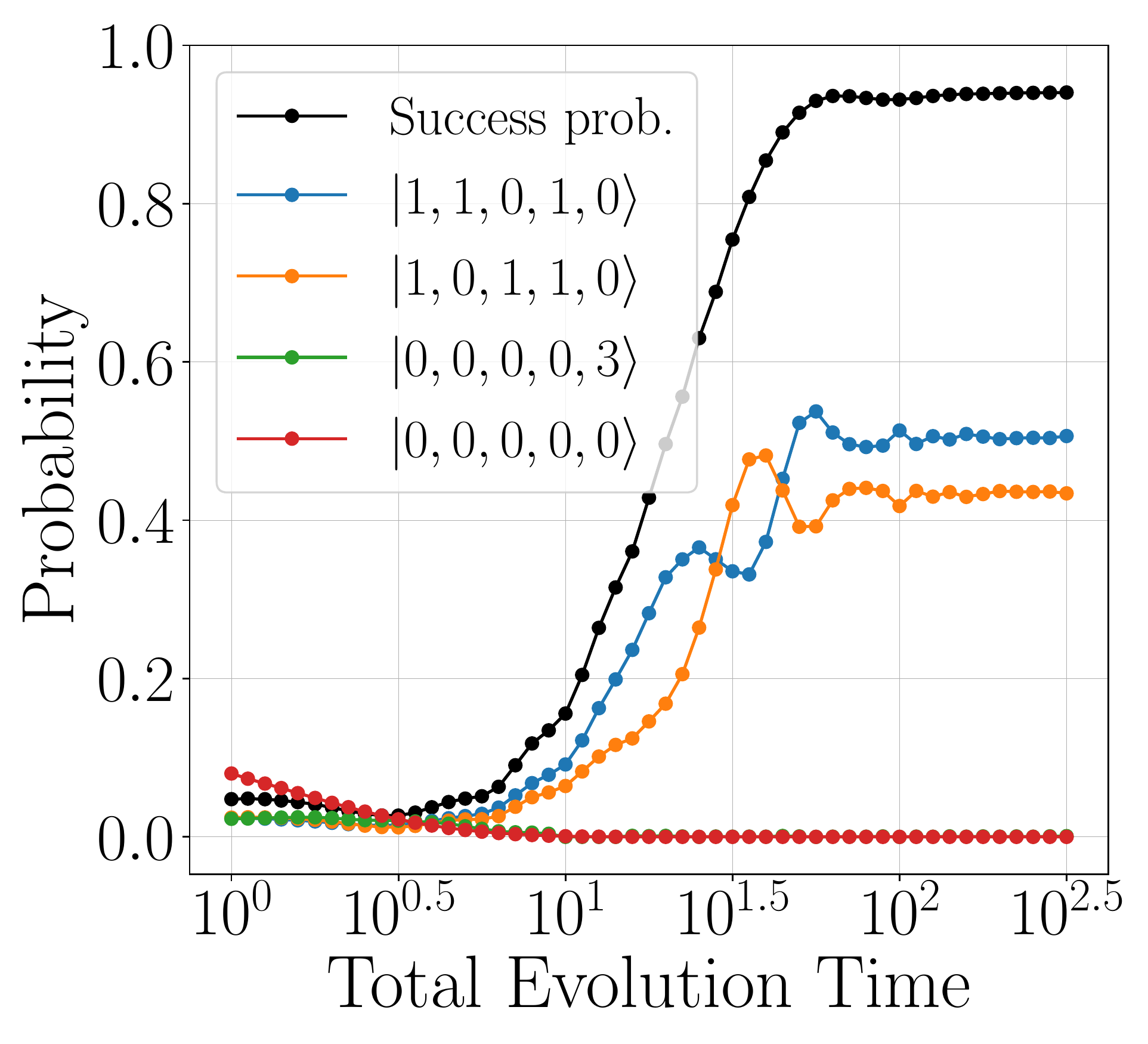}

}\subfloat[]{\includegraphics[scale=0.3]{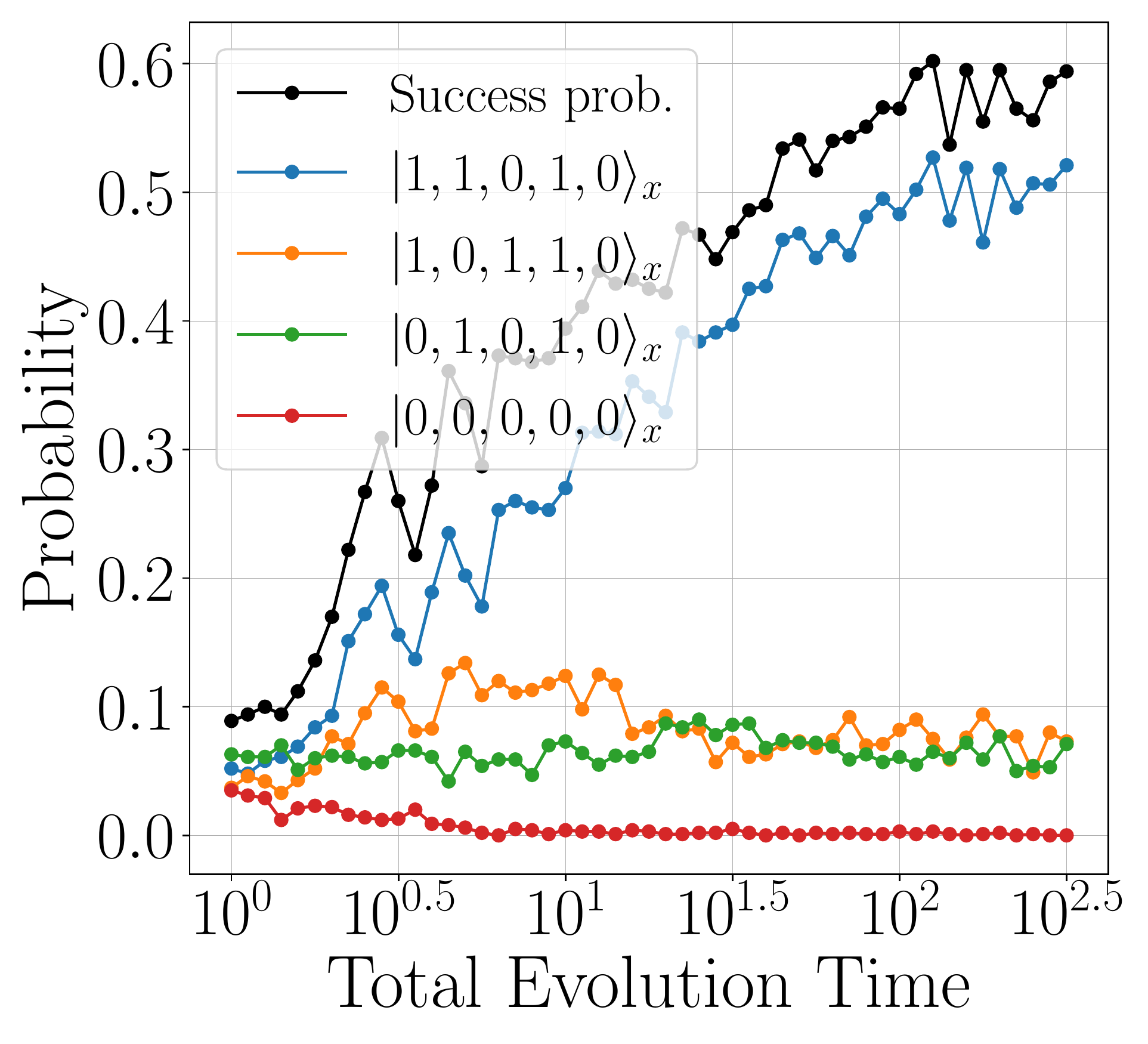}

}\subfloat[]{\includegraphics[scale=0.3]{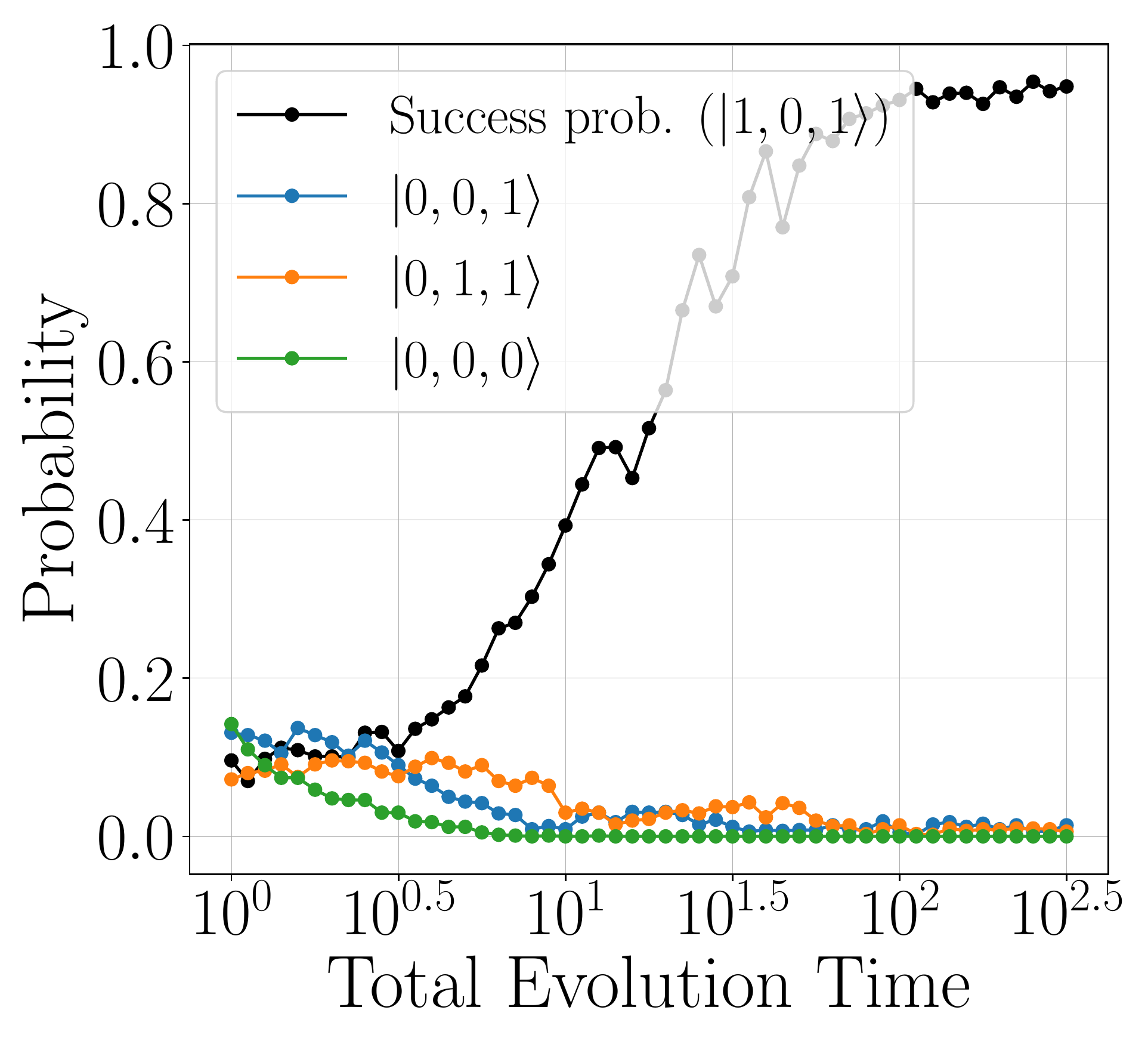}

}

\caption{Adiabaticity plots for the problem types studied: (a) simple integer linear programming (see~\cref{sec:ILP} of the main manuscript); \newline (b) the unbounded
knapsack problem (\cref{sec:ILP}); (c) the binary non-linear integer programming formulation of the \textsc{MaxClique} problem (\cref{sec:NLIP}); (d) the integer Motzkin--Straus
formulation of the \textsc{MaxClique} problem (\cref{sec:Non-convex}); (e) the continuous Motzkin--Straus formulation of the
\textsc{MaxClique} problem (\cref{sec:Non-convex}); and (f) the MIP sparse optimization problem (\cref{sec:MIP}). The plots show the evolution of the success probabilities of solving the problems, as well
as the probabilities of finding suboptimal
solutions, for different values of the total evolution time $T$. Note that here, for each value of $T$, each data point represents the probabilities at the end of an individual adiabatic evolution process, whereas in~\cref{fig:evolutions_IP} to \cref{fig:time-evolution_sparse_simplex} of the main manuscript, each plot represents a single evolution. Other than the parameter $T$,
the parameters used for these simulations are the same
as those used in~\cref{fig:evolutions_IP} to \cref{fig:time-evolution_sparse_simplex}. \label{fig:adiabaticity_plots}}
\end{figure*}

One common metric in studying the performance of the adiabatic quantum computation (AQC)
algorithm is the \emph{adiabaticity} of the method. This is done by studying the  performance of the algorithm with respect to different total evolution time values $T$. \Cref{fig:adiabaticity_plots} shows the evolution of the success
probabilities of solving the problem, as well as the probabilities of finding suboptimal solutions, for the integer and mixed-integer programming (MIP) problems discussed
in the main manuscript. As expected, the solutions for all of the problems show the similar trend
of starting at low success probabilities for a small total evolution
time $T$, and plateauing to their final success probability values for
larger values of $T$. These results indicate the optimum total evolution time for a given problem.

\section{Fair Sampling}
\label{sec:fair_sampling}

In this section, we look at fair sampling of the degenerate ground states of the integer programming
problems studied. In many applications, it is undesirable that the optimizer gives preference to one or a few of the solutions~\citep{konz2019uncertain} instead sampling all optimal solutions uniformly. For
instance, in the example graph instance shown in~\cref{fig:MaxClique_Example},
fair sampling in the \textsc{MaxClique} problem is attained when both maximum cliques $\{1,2,4\}$
and $\{1,3,4\}$ are found with equal probabilities by the solver.

\begin{figure}[!b]
\subfloat[\label{fig:bias_IP}]{\includegraphics[scale=0.3]{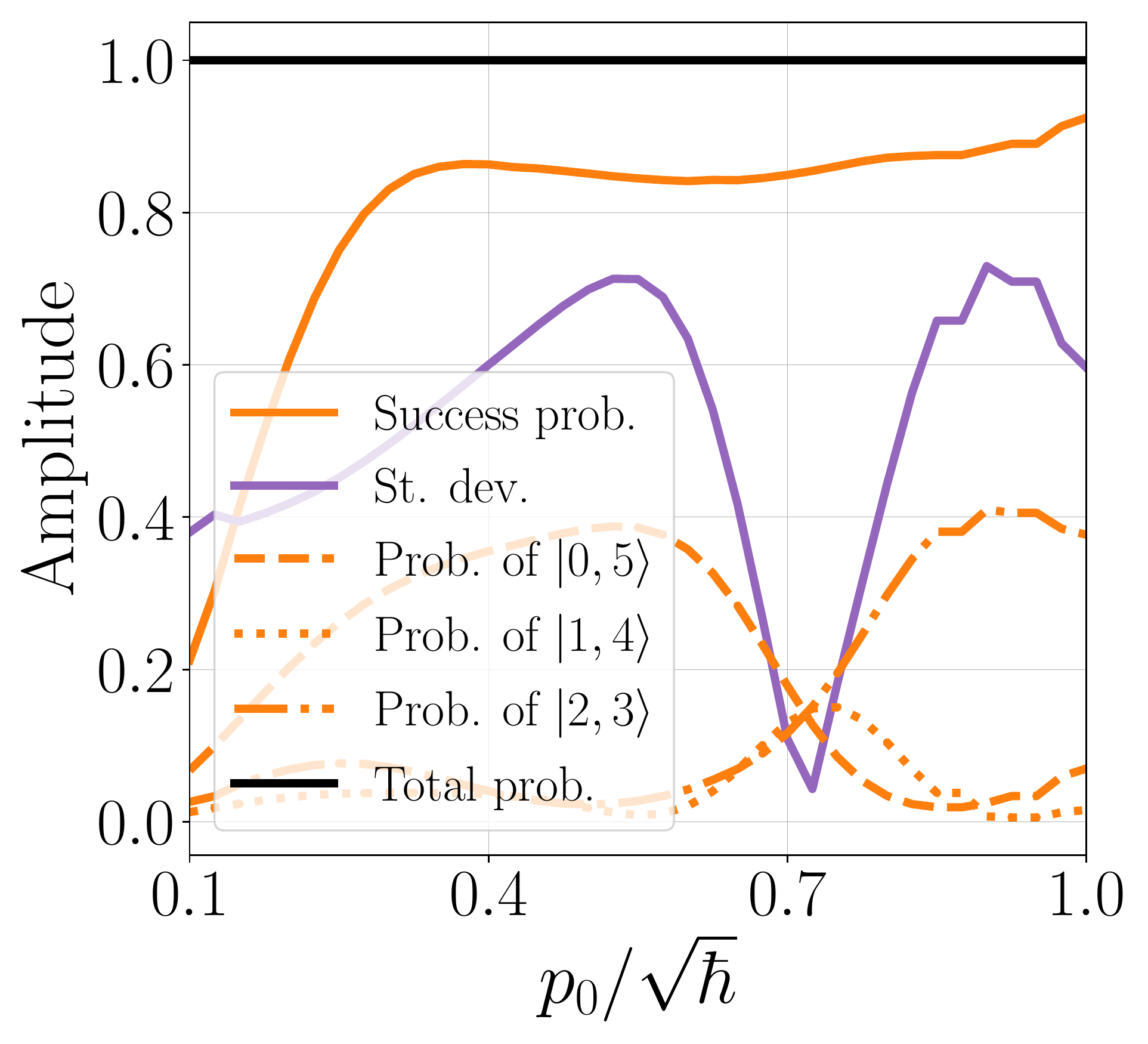}

}\subfloat[\label{fig:bias_MC-Motzkin}]{\includegraphics[scale=0.3]{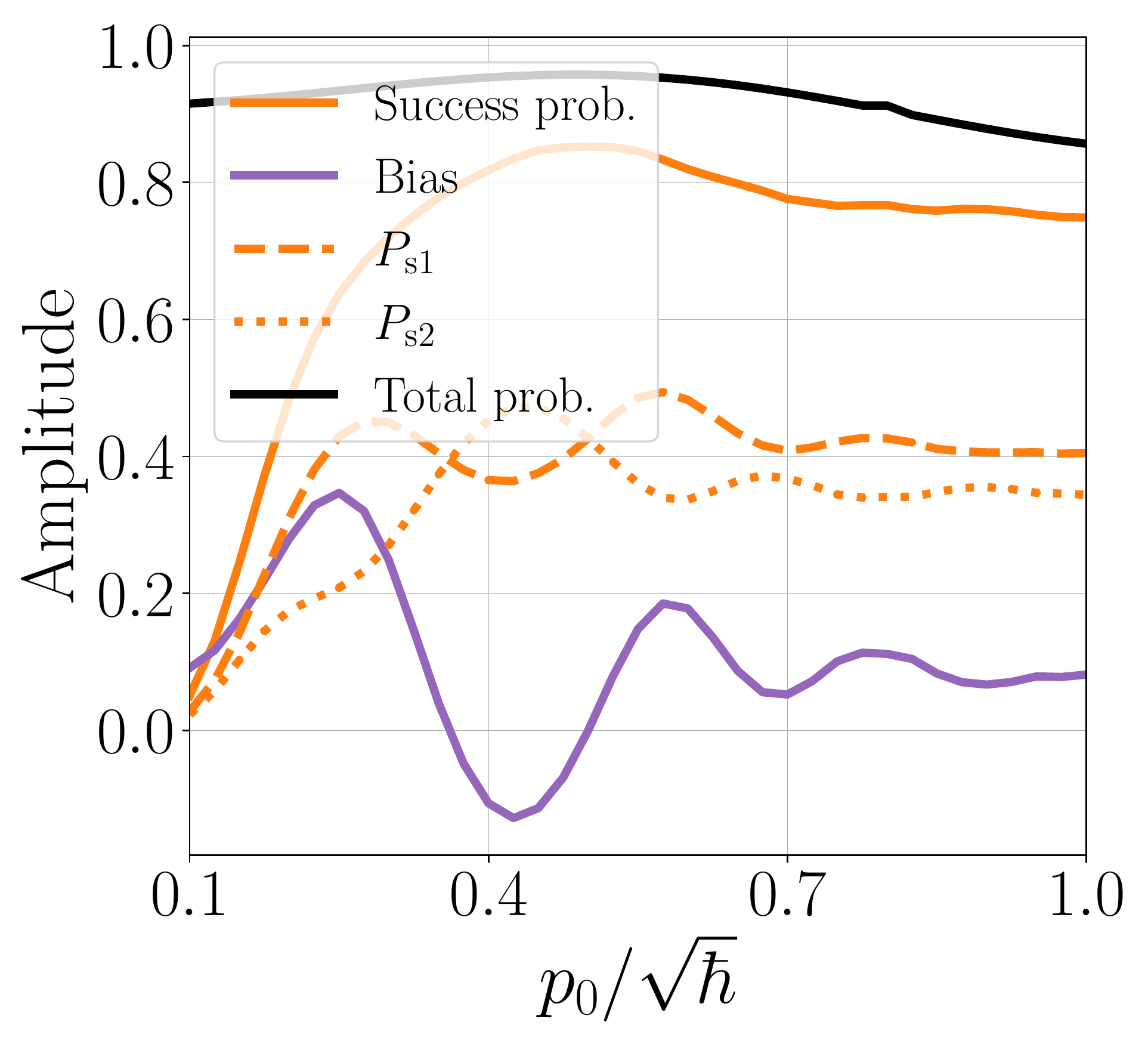}

}\subfloat[\label{fig:bias_MC-QUBO}]{\includegraphics[scale=0.3]{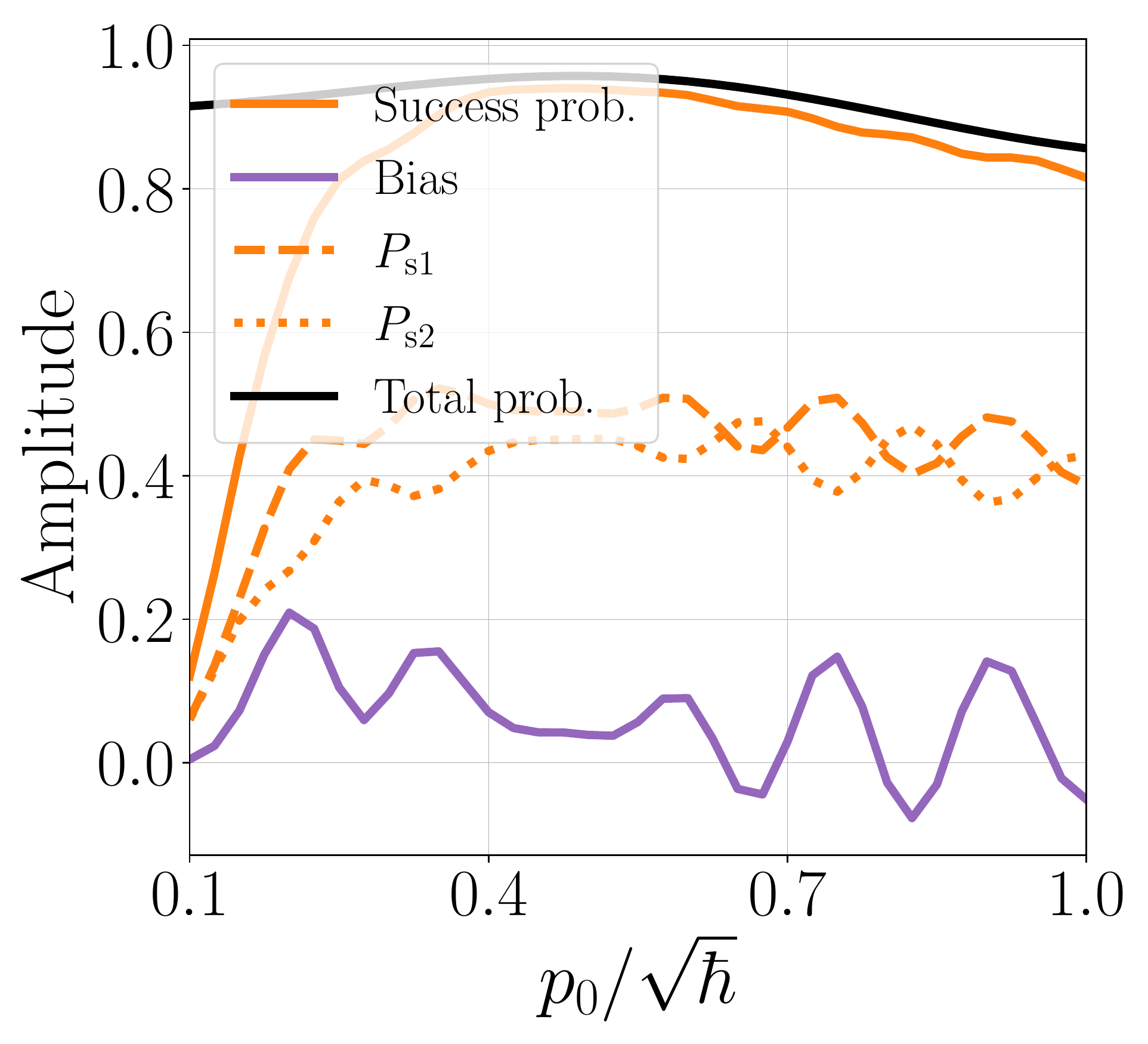}

}

\caption{Fair sampling results for the integer linear programming feasibility problem, 
and for the two non-integer linear programming formulations of the \textsc{MaxClique} problem. The success probabilities, and standard deviation or bias of the probabilities of the solution states, as a function of $p_{0}$, are shown using a solid orange and solid purple curve, respectively. All qumodes are initialized with $p_{0i}=p_{0}$. The bias for the integer linear programming problem is evaluated by finding the standard deviation of the probabilities
of the solution states. The bias in panels (b) and (c) for the \textsc{MaxClique} problem is defined as $b=\frac{P_{\text{s}1}-P_{\text{s}2}}{P_{\text{s}1}+P_{\text{s}2}}$,
where $P_{\text{s}1}$ and $P_{\text{s}2}$ are the success probabilities for the
two degenerate solutions of the studied problem. Other than the $p_{0i}$ values, the parameters used for these simulations are the
same as those used in \cref{fig:evolutions_IP}, \cref{fig:evolution_MC-QUBO_adiabatic}, and \cref{fig:MS-discrete}. The total probability (black curves) is obtained as the sum of probabilities of all quantum states within the truncated Fock space (resulting in a total probability value smaller than $1$). 
\label{fig:fair_sampling}}
\end{figure}

\begin{figure*}
\subfloat[\label{fig:ILP-probs-p0.2}]{\includegraphics[scale=0.3]{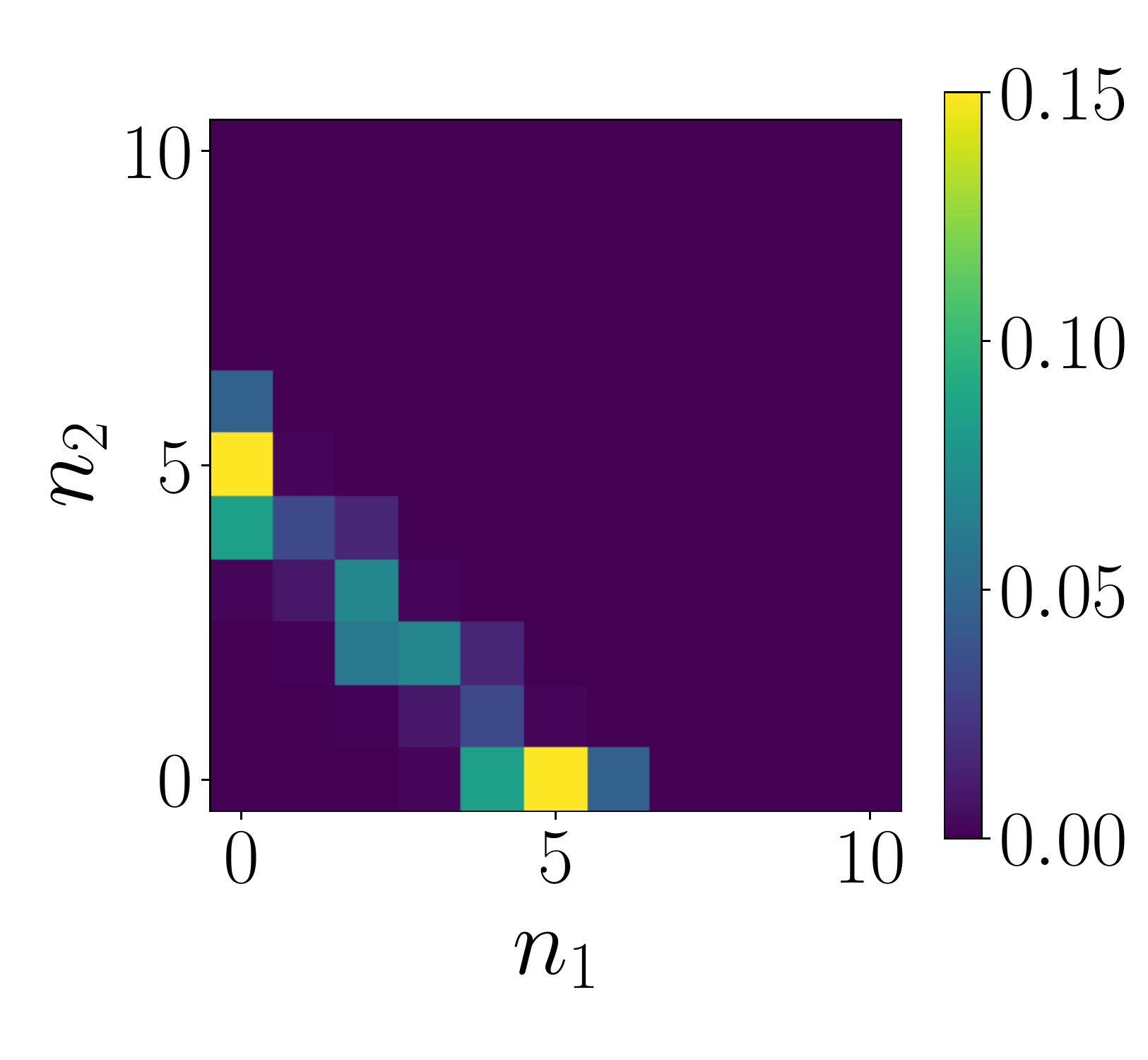}

}\subfloat[\label{fig:ILP-probs-p0.5}]{\includegraphics[scale=0.3]{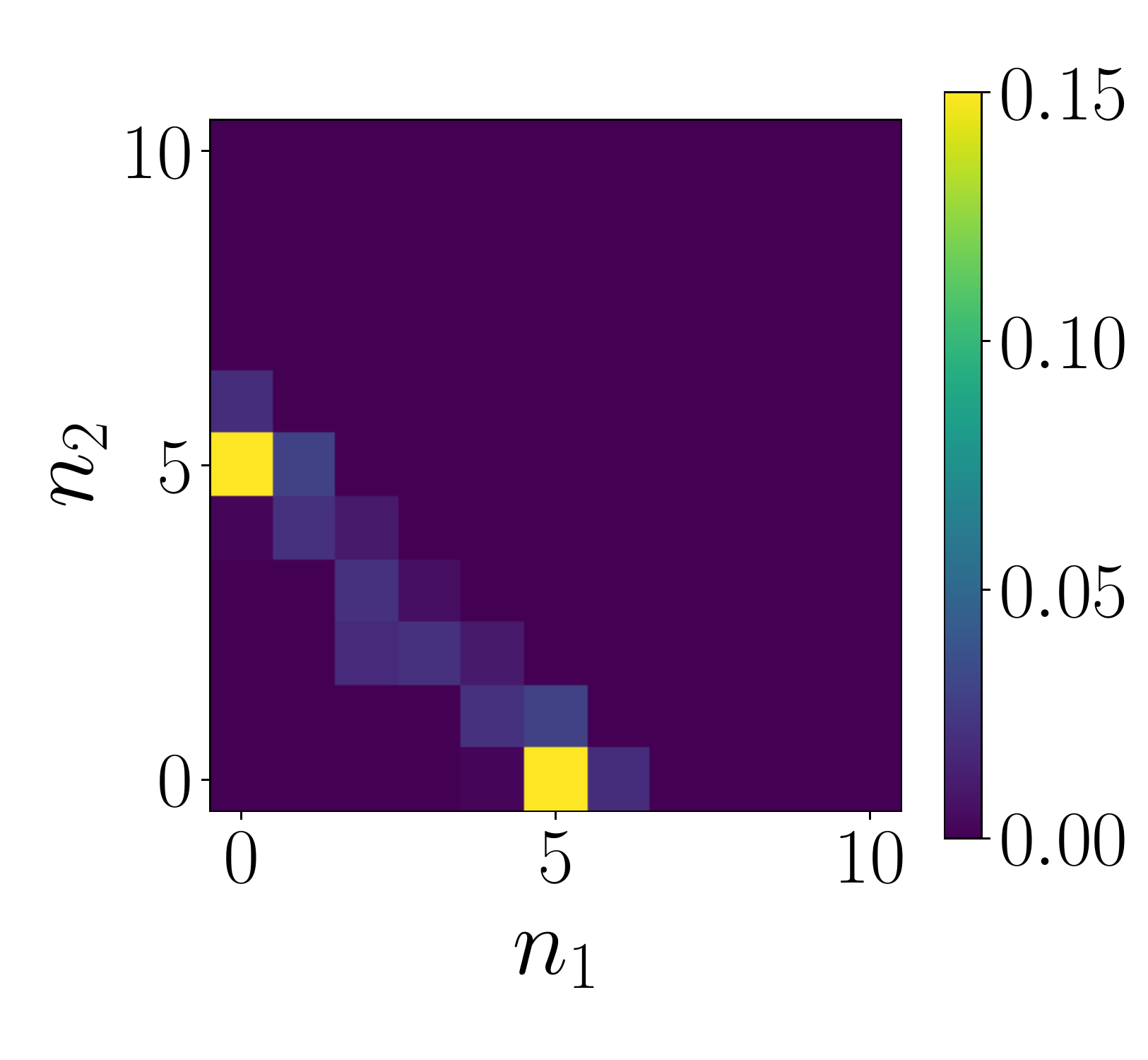}

}\subfloat[\label{fig:ILP-probs-p0.72}]{\includegraphics[scale=0.3]{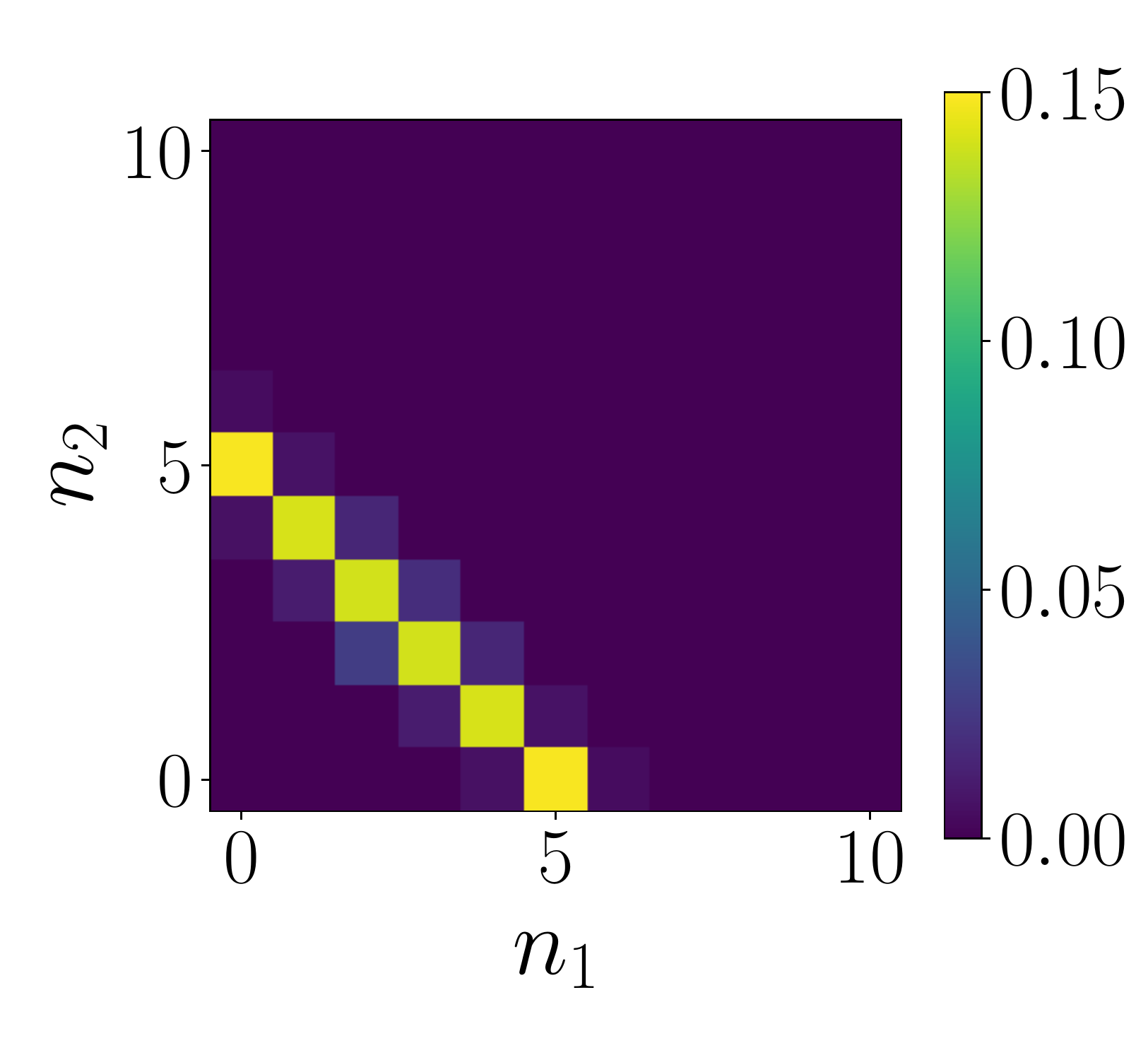}

}

\caption{Dependence of the probability distribution of the final quantum states, in the two-qumode Fock basis $\{\ket{n_1, n_2}\}$, on the parameter $p_0$ for the feasibility problem specified by the equation $n_1 + n_2 = 5$. The plots are for (a) $p_0/\sqrt{\hbar} = 0.2$, (b) $p_0/\sqrt{\hbar}=0.5$, and (c) $p_0/\sqrt{\hbar} = 0.72$. Note that the value $p_0 / \sqrt{\hbar} = 0.72$ corresponds to the most ``fair'' sampling of the six solution quantum states. This is also evident in~\cref{fig:bias_IP}, from the minimum of the standard deviation of the probabilities of the solution Fock states (shown using a purple curve), at $p_0/\sqrt{\hbar} = 0.72$.
\label{fig:ILP-biases-probs}}
\end{figure*}

\Cref{fig:fair_sampling} shows the fair sampling in the solutions of the integer linear programming (ILP) and \textsc{MaxClique} problems discussed in
the main manuscript. The plots show the biases of the probabilities of finding the
solutions as a function of the value of $p_{0}$ defining the mixing Hamiltonian~\cref{eq:Mixing}. All of the qumodes are initialized as the squeezed coherent states $|\boldsymbol{\alpha},-\boldsymbol{r}\rangle$, where $\boldsymbol{\alpha} = (\alpha_1, \alpha_2, ..., \alpha_N)$, with
$\alpha_j =ip_{0}/\sqrt{2}$ for all $j$. Having six degenerate solutions, the bias in the solutions of the  ILP feasibility problem $n_1 + n_2 = 5$ is described by the standard deviation of the final probabilities
of the solutions states $|0,5\rangle$, $|1,4\rangle$, $|2,3\rangle$,
$|3,2\rangle$, $|4,1\rangle$, and $|5,0\rangle$. A zero value for the standard deviation means that all six states have equal final
probabilities. \Cref{fig:bias_IP} shows the standard deviation
of the probabilities of the six solution states (shown using a purple curve), as well as
the probabilities of the three states $|0,5\rangle$, $|1,4\rangle$,
and $|2,3\rangle$, the success probability (the sum of the probabilities
of these six states), and the total probability (black curve). Due to
the symmetry of the problem between $n_1$ and $n_2$ in $\ket{n_1, n_2}$, the probabilities of finding the other three
states are the same as those for the states shown, with the boson number observables swapped. \Cref{fig:ILP-biases-probs} shows the probability distributions of the final quantum states in the two-qumode Fock basis $\{\ket{n_1, n_2}\}$, for three different $p_0$ values. As shown, by changing a single parameter $p_0$, one is able to adjust the distribution over the probabilities of the solution states, until a ``fair'' distribution has been reached. Here, a value of $p_0/\sqrt{\hbar} = 0.72$ gives the most even probability distribution among all solution quantum states for the studied ILP feasibility problem. This is also evident from the minimum of the standard deviation of these probabilities at $p_0 / \sqrt{\hbar} = 0.72$ in \cref{fig:bias_IP}.

\Cref{fig:bias_MC-Motzkin} and \cref{fig:bias_MC-QUBO} show the results of our fair sampling analysis for the \textsc{MaxClique} problem using formulations~\eqref{eq:MaxClique-Binary} and~\eqref{eq:motzkin_integer}. The bias value is defined as $b=\frac{P_{\text{s}1}-P_{\text{s}2}}{P_{\text{s}1}+P_{\text{s}2}}$,
where $P_{\text{s}1}$ and $P_{\text{s}2}$ are the final probabilities of finding the two
degenerate solutions $\{1,2,4\}$ and $\{1,3,4\}$, respectively.
As shown using a purple curve in \cref{fig:bias_MC-Motzkin} and \ref{fig:bias_MC-QUBO},
a bias of zero means that the two solutions have equal final probabilities. 

\section{Parameters in the Motzkin--Straus Formalism of the \textsc{MaxClique} Problem}
\label{sec:MS-hyperparam}

As mentioned earlier, for the integer representation of the Motzkin--Straus formalism~\eqref{eq:motzkin_integer}, the parameter
$\sigma$ is set equal to the size $\xi$ of the maximum clique(s) to find a solution
with high probability. As this value is not a priori known, a binary search can be performed to find its value by solving $\mathcal{O}(\log (|V|))$ instances of problem~\eqref{eq:motzkin_integer} using different choices of $\sigma \in \{1, \ldots, |V|\}$. In this section,
we study the case when the parameter $\sigma$ is larger than the size of the maximum clique of the problem and show that  even though optimization problem~\eqref{eq:motzkin_integer} may no longer be equivalent to problem~\eqref{eq:motzkin_continuous}, the correct results can still be inferred. 

\begin{figure}
\subfloat[]{\includegraphics[scale=0.3]{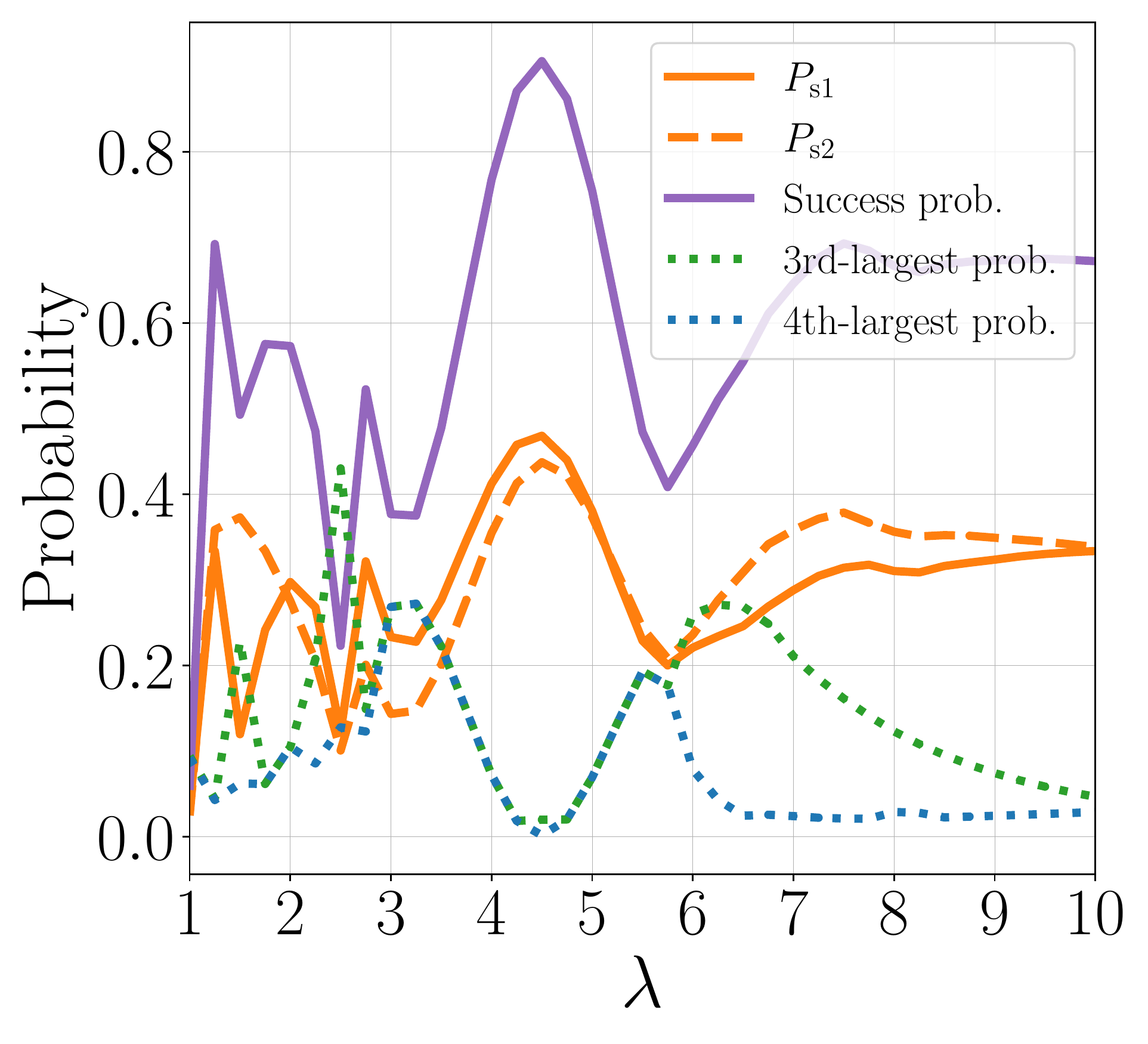}

}\subfloat[]{\includegraphics[scale=0.3]{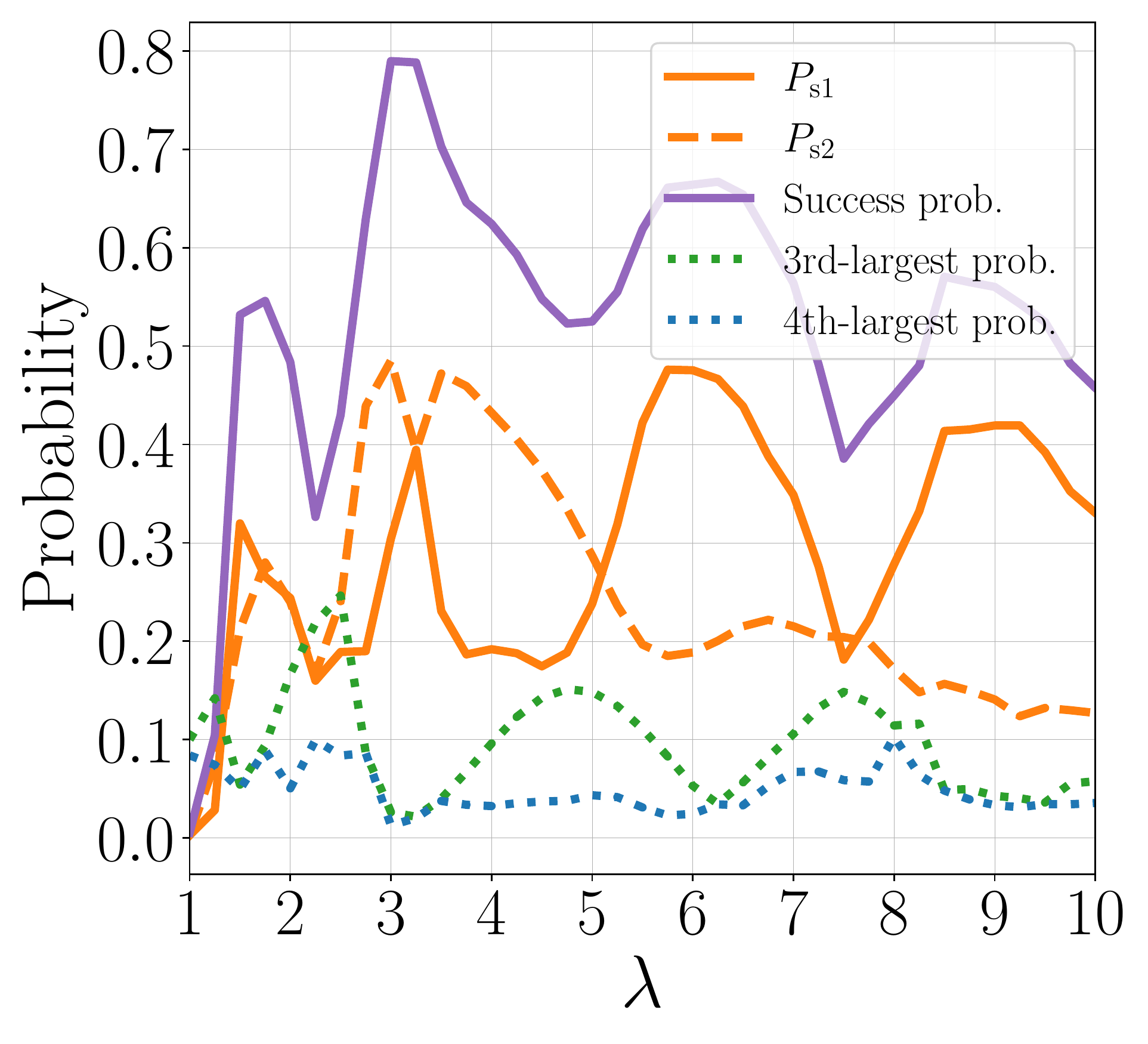}

}

\caption{Success probability of solving optimization problem~\eqref{eq:motzkin_integer} without knowing the maximum clique size $\xi$. Shown is the probability of finding the solutions $P_{\text{s}1}$ and $P_{\text{s}2}$ (defined in Eqs.~\eqref{eq:Ps1_=000026_Ps2}), as well as the success probability
$P_{\text{s}}=P_{\text{s}1}+P_{\text{s}2}$, when
$\sigma > \xi$, as a function of the hyperparameter $\lambda$ in the problem Hamiltonian~\eqref{eq:target_MS_discrete}. The results are for (a) $\sigma=4$ and (b) $\sigma=5$ for
the problem shown in \cref{fig:MaxClique_Example}, for which the maximum clique size is $\xi = 3$. As shown in these plots, by tuning $\lambda$, one is still able to find the solutions to the \textsc{MaxClique} problem. 
\label{fig:figure_sucess_P_W}}
\end{figure}

\Cref{fig:figure_sucess_P_W} shows the success probability,
as well as the probability of finding either of the two solutions,
as a function of the hyperparameter $\lambda$ for the \textsc{MaxClique} problem shown in \cref{fig:MaxClique_Example}. The probability of finding a particular solution here
is defined as the sum of the probabilities of all of the states that
have one or more photons in the qumodes corresponding to the maximum
clique. For the maximum cliques given by the vertices $\{1,2,4\}$
and $\{1,3,4\}$, the solution probabilities
are given by
\begin{equation}
P_{\text{s}1}=\sum_{n_{1},n_{2},n_{4}=1}^{d}\langle n_{1},n_{2},0,n_{4},0|\psi\rangle \qquad \text{and} \qquad P_{\text{s}2}=\sum_{n_{1},n_{3},n_{4}=1}^{d}\langle n_{1},0,n_{3},n_{4},0|\psi\rangle,
\label{eq:Ps1_=000026_Ps2}
\end{equation}
respectively, where $\ket{\psi}$ is the final quantum state prepared using AQC. Here, $d$ is the truncation dimension of the Fock space, which is physically infinite but in a classical simulation must be finite. As shown in~\cref{fig:figure_sucess_P_W}, there is an optimal choice of $\lambda$ at which the largest success probability for a given value of $\sigma$ is reached. This
method of finding the success probabilities aligns well with the functionality of
single-photon avalanche detectors (SPAD), where the arrival
of one or more photons is detected without the capability to resolve the photon number. The results in \cref{fig:figure_sucess_P_W} indicate that the maximum cliques correspond to the
qumodes that have nonzero photon numbers and thus can be measured
using currently available SPADs.

\clearpage
\bibliography{refs}

\end{document}